\documentstyle[11pt,epsf,dina4p]{article}
\begin{document}
\newcommand{\figXa}     {running_At2}         % running of At
\newcommand{\mhmt}      {mh_mt}               % light CP even Higgs mass m_h
\newcommand{\figII}     {hz0_low_high}  % light CP even Higgs mass m_h
\newcommand{\figIII}    {ma}                  % CP odd Higgs mass m_A
\newcommand{\mze}  {{\ifmmode m_0           \else $m_0$            \fi}}
\newcommand{\mha}  {{\ifmmode m_{1/2}       \else $m_{1/2}$        \fi}}
\newcommand{\tb}   {{\ifmmode \tan\beta     \else $\tan\beta$      \fi}}
\newcommand{\mz}   {{\ifmmode M_{Z}         \else $M_{Z}$          \fi}}

\newcommand{\beq}   {\begin{eqnarray}}
\newcommand{\eeq}   {\end{eqnarray}}
\newcommand{\ba}    {\begin{array}}
\newcommand{\ea}    {\end{array}}
\newcommand{\nn}   {\nonumber \\}
\newcommand{\smas}[2]{\tilde{m}^#2_{#1}}

\newcommand{\rb}[1]{\raisebox{1.5ex}[-1.5ex]{#1}}\begin{titlepage}

\renewcommand{\floatpagefraction}{0.6}
\renewcommand{\textfraction}{0.2}
\begin{flushright}
\vspace*{-2.2cm}
\noindent
IEKP-KA/96-03     \\
JINR E2-95-401    \\
hep-ph/9603346    \\
March 12th, 1996  \\
\end{flushright}
\vspace{1.7cm}

\begin{center}
{\Large\bf MSSM Predictions of the} \vspace{4pt}\\
{\Large\bf Neutral Higgs Boson Masses}\vspace{4pt}\\
{\Large\bf and  LEP II Production Cross Sections}\vspace{4pt}\\
{\bf A.V.~Gladyshev\footnote{E-mail: gladysh@thsun1.jinr.dubna.su},
D.I.~Kazakov\footnote{E-mail: kazakovd@thsun1.jinr.dubna.su}}       \\
{\it Bogoliubov Lab. of Theor. Physics,
     Joint Inst. for Nucl. Research,}                             \\
{\it 141 980 Dubna, Moscow Region, Russia}                          \\
and                                                                 \\
{\bf W.~de Boer\footnote{E-mail: DEBOERW@CERNVM},
G.~Burkart\footnote{E-mail: gerd@ekpux3.physik.uni-karlsruhe.de},
R.~Ehret\footnote{E-mail: ralf.ehret@cern.ch}}                     \\
{\it Inst.\ f\"ur Experimentelle Kernphysik, Univ.\ of Karlsruhe,}   \\
{\it Postfach 6980, D-76128 Karlsruhe, Germany}                   \\
\end{center}
\vspace{1cm}

\begin{center}
{\bf Abstract}
\end{center}

\parindent0.0pt
Within the framework of the Minimal Supersymmetric
Standard Model (MSSM) the Higgs masses and LEP II production cross
sections are calculated for a wide range of the parameter space.  In
addition, the parameter space restricted by unification, electroweak
symmetry breaking and other low energy constraints is considered in
detail, in which case the masses of all SUSY partners can be
estimated, so that their contributions to the radiative corrections
can be calculated. Explicit analytical formulae for these
contributions are derived.  The radiative corrections from the Yukawa
couplings of the third generation are found to dominate over the
contributions from charginos and neutralinos. Large Higgs mass
uncertainties are due to the top mass uncertainty and the unknown sign
of the Higgs mixing parameter. For the low $\tb$ scenario the mass of
the lightest Higgs is found to be below 90 GeV for a top mass below
180 GeV.  The cross section at a LEP II energy of 192 GeV is
sufficient to find or exclude this  scenario.
For the high $\tb$ scenario only a small fraction of
the parameter space can be covered, since the Higgs mass is predicted
between 105 and 125 GeV in most cases. At the theoretically possible
LEP II energy of 205 GeV part of the parameter space for the large
$\tb$ scenario would be accessible.

\end{titlepage}

\setcounter{page}{1}
\section{Introduction}

In recent years the Minimal Supersymmetric Standard Model
(MSSM)~\cite{MSSM} has become the subject of intensive investigation.
It is the simplest extension of the Standard Model (SM) which
involves supersymmetry and provides us with a promising
candidate for a Grand Unified Theory~\cite{ABF}.

One of the striking features of the MSSM is the strict prediction
of  the lightest Higgs boson mass, since the Higgs coupling constant is
not an arbitrary parameter, but   fixed by the gauge couplings leading
to a tree-level mass less than $M_Z$:  $ m_h \le M_Z.  $

Since Higgs masses up to $\mz$ will be observable at LEP II after its
energy upgrade to 192 GeV, the prediction of the Higgs mass
spectrum and calculation of the cross sections has been the subject of
large working groups~\cite{lepwg}.  If the radiative corrections are
included, one obtains an upper limit of about 150 GeV for the
lightest CP-even Higgs mass~\cite{Ellis,other}.

A more precise prediction can be obtained, if one considers
simultaneously the constraints for which the constrained MSSM (CMSSM)
has become so attractive, namely the unification of the three gauge
couplings, radiative electroweak symmetry breaking (EWSB) at the $M_Z$
scale and unification of the Yukawa couplings~\cite{Many,WE,WE1}.

Assuming soft supersymmetry breaking at the GUT scale, all the masses
of the super partners and the Higgs bosons can be expressed in terms of
five free parameters~\cite{soft}. Strong constraints and correlations
between these parameters can be obtained from fits to low energy data.
In this case the loop corrections to the Higgs masses from all SUSY
particles can be calculated. 

It is the purpose of this paper to present the complete set of
analytical expressions of the one-loop radiative corrections to the
effective potential giving rise to the masses of the neutral Higgs
bosons, including the $D$-terms proportional to the electroweak
gauge couplings, and taking into account the contributions
from all particles of the MSSM.  The masses are calculated using the effective potential
approach (EPA) and will be compared with the full one-loop diagrammatic
calculations (FDC)~\cite{Bri,CPR}, and the dominant second order
calculations~\cite{cw}.  For the Higgs-strahlung the corresponding
cross sections at LEP II  will be calculated.  Within the CMSSM the
associated pair production $e^+e^-\rightarrow hA$ is negligible at LEP
II, as will be shown below.

\section{One-loop effective Higgs potential}

The one-loop  effective Higgs potential in the MSSM
can be written as~\cite{Arnowitt}:
\begin{eqnarray}
V &=& V_{tree}+\Delta V \nonumber \\
V_{tree}&=&m_1^2|H_1|^2+m_2^2|H_2|^2-m_3^2(H_1H_2+h.c.)
+\frac{g^2+g'^2}{8}
(|H_1|^2-|H_2|^2)^2+\frac{g^2}{2}|H_1^*H_2|^2  \nonumber \\
\Delta V &=&\sum_k \frac{1}{64\pi^2}(-1)^{2J_k}(2J_k+1)c_k
m_k^4 \left(\log\frac{m_k^2}{Q^2}-\frac{3}{2}\right),
\end{eqnarray}
where the sum is taken over all particles in the loop;
$c_k=c_{colour}c_{charge}$ with $c_{colour}=3(1)$
for coloured (uncoloured)
particles and $c_{charge}=2(1)$ for charged (neutral) particles,
$J_k$ is the spin  and $m_k$ are the field-dependent masses
of the particles in the loops at the scale $Q$.
Expressions for $m_k$ have been summarized in Appendix A.

Due to EWSB the Higgs doublets
$$ H_1 = \left(\begin{array}{c}H^0_1 \\
H^-_1\end{array}\right), \ \ \
 H_2 = \left(\begin{array}{c}H^+_2 \\
H^0_2\end{array}\right),$$
obtain non-zero vacuum expectation values $<H_1>=
\left(\begin{array}{c}v_1\\
0\end{array}\right), \  <H_2>=
\left(\begin{array}{c}0\\
v_2\end{array}\right)$, where $v_1=v\cos\beta, v_2=v\sin\beta$.

Minimization  of the potential with respect to the vacuum expectation
values of the Higgs fields leads to the following mimimization conditions:
\beq
  \frac{M_Z^2}{2} & = & \frac{(m_1^2+\Sigma_1)-(m_2^2+\Sigma_2)
    \tan^2\beta}{\tan^2\beta-1},                         \label{min1} \\
   2m_3^2         & = &\sin 2\beta (m_1^2+m_2^2+\Sigma_1+\Sigma_2),
                                                          \label{min2}
\eeq
where
\beq
  \Sigma_1 & = &
    \frac{1}{64\pi^2} \sum_k (-1)^{2J_k}(2J_k+1)c_k\frac{1}{v_1}
    \frac{\partial m_k^2}{\partial v_1}f(m_k^2) ,  \label{Sigma1}\\
  \Sigma_2 & = &
    \frac{1}{64\pi^2} \sum_k (-1)^{2J_k}(2J_k+1)c_k\frac{1}{v_2}
    \frac{\partial m_k^2}{\partial v_2}f(m_k^2) .  \label{Sigma2}
\eeq
The function $f(m_k^2)$ is defined as:
\beq
    f(m_k^2) & = & m_k^2 \left(\ln\frac{m_k^2}{Q^2}-1\right).
    \label{fmi2}
\eeq

Expressions for all contributions to the one-loop corrections
$\Sigma_1$ and $\Sigma_2$  are given in Appendix B.

\section{Corrections to the Higgs Masses}
At tree level the CP-even Higgs masses are given by:
\begin{eqnarray}
m^2_{H,h}&=&\frac{1}{2}\left[m_A^2+M_Z^2 \pm \sqrt{(m_A^2+M_Z^2)^2
- 4m_A^2M_Z^2\cos^2 2\beta }\right]\;. \label{mh}
\end{eqnarray}
Usually one considers the mass of the CP-odd Higgs $m_A$ and
$\tan\beta$ as the two independent parameters of $m_h$ and $m_H$.
However, the radiative corrections depend on the masses of all
particles running inside the loops. These would represent   many
additional free parameters, if all considered independent. Therefore we
will examine the supergravity inspired MSSM, in which all particle
masses depend on 5 parameters at the GUT scale: the common breaking
mass $m_0$ for all spin 0 sparticles (squarks and sleptons), the common
breaking mass $m_{1/2}$ for the spin 1/2 gauginos, the Higgs mixing
parameter $\mu$, the bilinear Higgs coupling $B$ and the trilinear
couplings $A_t^0=A_b^0=A_\tau^0=A^0$. The bilinear coupling $B$
is related to $m_3$ through $m_3^2 = B\mze\mu$. Substituting this
relation into eq.~(\ref{min2}) the bilinear coupling $B$ can be
expressed through the ratio $v_2/v_1\equiv \tan\beta$ of the two vacuum
expectation values at the scale $M_Z$.  Within this framework all
one-loop corrections can be calculated.

For  given values of   $m_0$ and $m_{1/2}$  the parameters related to
the Higgs sector ($\mu, \tan\beta, A^0$) are strongly constrained by
EWSB and bottom-tau unification~\cite{Many,WE,WE1}. The latter
constraint yields two preferred solutions for $m_t=179 \pm 12
\; GeV$~\cite{cdfd0}: the low
$\tan\beta$ scenario ($\tan\beta\approx 1.7$) which corresponds to a
top Yukawa coupling much larger than the bottom Yukawa coupling
$Y_t \gg Y_b$ and the high $\tan\beta$ scenario
($\tan\beta \approx 41$) with $Y_t\approx Y_b$.

%RE %DK new paragraph
%RE One should mention, however, that analytical formulae presented
%RE in this paper are obtained using the effective potential approach (EPA)
%RE which implies zero momentum contribution and ignores the contribution
%RE to the wave function renormalization. The difference between our
%RE method and, for example, the explicit diagrammatic
%RE calculations~\cite{Bri,CPR} can achieve several GeV, depending
%RE on the tree-level masses. But for the light Higgs boson EPA stays
%RE rather good approximation.

\paragraph{ CP-odd Higgs Mass $m_A$.} \  \

The masses of the CP-odd neutral Higgses are given by the
eigenvalues of the matrix
\begin{eqnarray}
{\cal M}\equiv\frac{1}{2}\left(\frac{\partial^2 V}
  {\partial \phi_i \partial \phi_j}
\right)_{H_i=v_i} & = &
\frac{1}{2} \left(\begin{array}{cc}
                     \tan\beta &     1    \\
                         1     & \cot\beta
                  \end{array}\right)
  (2m^2_3+\Delta),
\label{Mma}
\end{eqnarray}
where $m_3^2$ is given by eq.~(\ref{min2}) and $\phi_i$ are the
imaginary components of $H^0_i=v_i+\psi_i+i\phi_i$.  The mass
eigenstates are the massless Goldstone boson and
\begin{equation}
m^2_{A_{1-loop}}=m_1^2+m_2^2+\Sigma_1+\Sigma_2+\Delta /\sin2\beta=
\frac{ 2m^2_3+\Delta}{\sin2\beta},
\label{ma}
\end{equation}
where \begin{eqnarray} \Delta &=&\frac{g^2}{32\pi^2}\frac{\mu
m_0}{M_W^2} \sum_k c_k s_k A_k m_k^2 \frac{f(\tilde m^2_{k_1})-f(\tilde
m^2_{k_2})} {\tilde m^2_{k_1}-\tilde m^2_{k_2}} \nonumber  \\ &+&
\frac{g^2}{4\pi^2}M_2\mu\frac{f(\tilde m^2_{\chi_{1}^{\pm}})-f(\tilde
m^2_{\chi_{2}^{\pm}})}{\tilde m^2_{\chi_{1}^{\pm}}-\tilde
m^2_{\chi_{2}^{\pm}}} \label{macorr}              \\
&+&\frac{g^2+g'^2}{8\pi^2}\sum_{k=1}^4\frac{\mu\lambda_k(\lambda_k-
M_2\sin^2\theta _W-M_1\cos^2\theta _W)}
{D(\lambda_k)}f(\lambda_k^2). \nonumber
\end{eqnarray}
$M_1$ and $M_2$ are the masses of the $U(1)_Y$ (Bino)
and $SU(2)_L$ (Wino) gauginos,
respectively. In principle the sum in the first term is taken over
all sparticles, but since the contributions are proportional to the
Yukawa couplings, the first two generations are negligible and one
can take $k= t,b,\tau$; $c_k$ is the colour/charge factor
($c_k = 2$ for sleptons and $c_k = 6$ for squarks),
$s_k=\sin^{-2}\beta $ for up squarks and $s_k=\cos^{-2}\beta $ for
down squarks and sleptons; $\lambda_k$ are the neutralino masses,
which can be obtained from the
%RE numerical
solutions of the quartic equation $F(\lambda )=0$
defined in Appendix A, and $D(\lambda_k)$ is its derivative with
respect to $\lambda_k$. The function $f(\lambda^2_k)$ is defined
in eq.~(\ref{fmi2}).
The renormalization scale $Q$ is taken to be of the order of the
electroweak scale ($\mz$ or the pole mass of the top quark $M_t$).
\bigskip

%
%----------------- A_t running %
%\clearpage
\begin{figure}[th]
  \vspace*{-1.0cm}
  \begin{center}
    \leavevmode
    \epsfxsize=12cm
    \epsffile[0 65 675 420]{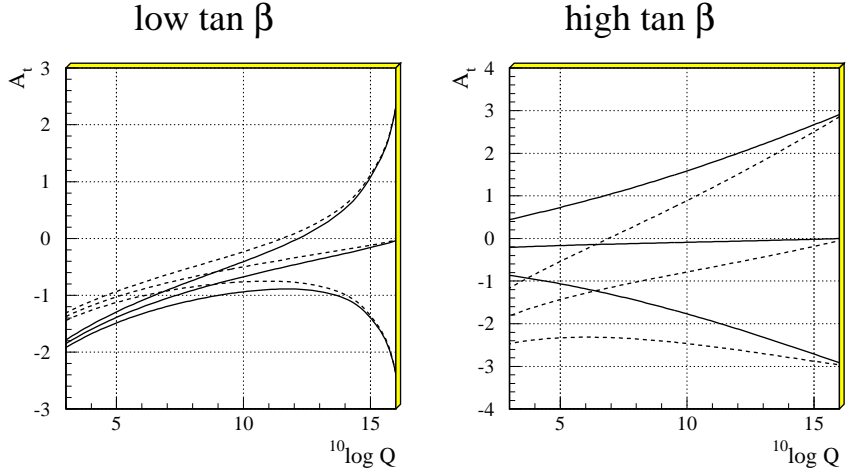}
    \vspace*{-0.5cm}
    \caption[]{\label{f1}The running of $A_t$.  The values of
      ($\mze,\mha$) were chosen to be (200,270) and (800,90) GeV
      for the low and high $\tb$ scenario, respectively (solid lines)
      and (1000,1000) (dashed lines). At low values of $\tb$ a
      strong convergence to a single value is found for all values
      of $A^0$  at the GUT scale, while for high values this tendency is
      less pronounced. The solid lines correspond to the preferred
      solutions of ref.~\cite{WE1}.}
  \end{center}
\end{figure}

\paragraph{ CP-even Higgs Masses $m_{h,H}$.} \  \

The masses of the CP-even neutral higgses $H$ and $h$ are determined
by the mass matrix
\beq
{\cal M}=\frac{1}{2}
\left(\frac{\partial^2 V}{\partial \psi_i \partial \psi_j}
\right)_{v_1,v_2}&=&
\left(\begin{array}{cc}
\tan\beta & -1 \\
-1 & \cot\beta
\end{array}\right) m_3^2 +\frac{1}{2}
\left(\begin{array}{cc}
\cot\beta & -1 \\
-1 & \tan\beta
\end{array}\right) M_Z^2\sin 2\beta   \nn
&+& \frac{1}{2}
\left(\begin{array}{cc}
\Delta_{11} & \Delta_{12}\\
\Delta_{12} & \Delta_{22} \label{Mmh}
\end{array}\right)
\eeq where $\psi_i$ are the real components of $H^0_i$
and
\beq
  \Delta_{11} & = & \frac{1}{32\pi^2}\sum_k (-1)^{2J_k}(2J_k+1)c_k\left[
                    \left(\frac{\partial m_k^2}{\partial \psi_1}\right)^2
                    \log\frac{m_k^2}{Q^2}+\left(\frac{\partial^2 m_k^2}
                    {\partial \psi_1^2} -\frac{1}{\psi_1}\frac{\partial m_k^2}
                    {\partial\psi_1}\right) f(m_k^2)\right]                \nn
  \Delta_{22} & = & \frac{1}{32\pi^2}\sum_k (-1)^{2J_k}(2J_k+1)c_k
                    \left[\left(\frac{\partial m_k^2}{\partial \psi_2}\right)^2
                    \log\frac{m_k^2}{Q^2}+\left(\frac{\partial^2 m_k^2}
                    {\partial \psi_2^2} -\frac{1}{\psi_2}\frac{\partial m_k^2}
                    {\partial\psi_2}\right) f(m_k^2)\right]                \nn
  \Delta_{12} & = & \frac{1}{32\pi^2}\sum_k(-1)^{2J_k}(2J_k+1)c_k\left[
                    \frac{\partial m_k^2}{\partial \psi_1}\frac{\partial m_k^2}
                    {\partial \psi_2}\log\frac{m_k^2}{Q^2}+
                    \frac{\partial^2 m_k^2}
                    {\partial \psi_1 \partial \psi_2}f(m_k^2)\right].
                                                                \label{higgs1}
\eeq
Diagonalization of the mass matrix
yields after substituting the minimization conditions
  (\ref{min1}) and (\ref{min2}):
\newcommand{\mdrei}{\frac{2m_3^2}{\sin 2\beta}}
\beq
  m^2_{h,H}
    & = & \frac{1}{2}\left[\mdrei+M^2_Z +\Delta_{11}+\Delta_{22}
          \mp\sqrt{\Delta^2_{h,H}}\right] \label{higgs}\\
          \Delta^2_{h,H}&=&(\mdrei+M_Z^2+\Delta_{11}+\Delta_{22})^2  \nn
    & - & 4\mdrei M_Z^2\cos^22\beta-4(\Delta_{11}\Delta_{22}
          -\Delta_{12}^2) \nn
    & - & 4(\cos^2\beta M^2_Z+\sin^2\beta \mdrei)\Delta_{22}         \nn
    & - & 4(\sin^2\beta M^2_Z+\cos^2\beta \mdrei)\Delta_{11}         \nn
    & - & 4\sin2\beta (M^2_Z+\mdrei)\Delta_{12}. \label{CPevenmas}
\eeq
 The expressions for
  $\Delta_{ij}$ have been evaluated in Appendix C.

If $m_A \gg M_Z$ and if one considers only the contributions from the
(s)top sector the formula for the lightest Higgs can be approximated
as \cite{cw}:
\begin{eqnarray}
  m^2_{h} & = & M_Z^2\cos^2 2\beta
  \left(1-\frac{3}{8\pi^2}\frac{m_t^2}{v^2} t\right) \label{higgs2} \\
          & + & \frac{3}{4\pi^2}\frac{m_t^4}{v^2}
          \left[\frac{1}{2}\tilde{X}_t+t+\frac{1}{16\pi^2}
            \left(\frac{3}{2}\frac{m_t^2}{v^2}-
              32\pi\alpha_s(M_t)
            \right)
            (\tilde{X}_t+t^2)
          \right] \nonumber
\end{eqnarray}
where
$\tilde{X}_t=2\tilde{A}_t^2/M^2_{SUSY}(1-\tilde{A}_t^2/12M^2_{SUSY})$,
$\tilde{A}_t=A_t\mze-\mu\cot\beta,$
$t=\log(M^2_{SUSY}/M_t^2)$, $m_t=m_t(M_t)$ and
$v=174$ GeV.
$M_{SUSY}$ is a typical scale of the squark masses, which can be
defined as the arithmetic average of the stop squared mass eigenvalues
for not too large  stop mass splittings. The last terms in
eq.~(\ref{higgs2}) proportional to $m_t^6$ and the strong
coupling constant $\alpha_s$ take into account the most relevant
%RE two-loop corrections (which are absent in eq.~\ref{CPevenmas}).
two-loop corrections, which are absent in eq.~(\ref{CPevenmas}).

It is important to note that the value of the trilinear coupling $A_t$
at the electroweak scale is not a free parameter,
at least for low values of $\tb$. In that case $A_t(M_Z)$ goes to a
fixed point solution, i.e.~its value becomes independent of the
starting value $A^0$  at the GUT scale. This is demonstrated
in fig.~\ref{f1} for several values of $m_0$ and $m_{1/2}$
and can be easily understood for the solution of the RGE's in
case $Y_b \ll Y_t$:
\begin{equation}
A_t(M_Z)= A^0\left(1-\frac{Y_t}{Y_f}\right)-
\frac{m_{1/2}}{m_0}\left(3.6-2.1\frac{Y_t}{Y_f}\right)
\approx 0.009A^0-1.7\frac{m_{1/2}}{m_0}, \label{aa}
\end{equation}
where $Y_f$ is the fixed point solution of the Yukawa RGE and
the numerical coefficients are taken from ref.~\cite{WE}.
For $m_{1/2}>m_0 /6$ the last term is dominating, as long as $|A^0|<3$
(in units of $m_0$), which is the typical range for which tachyonic
solutions and colour breaking minima in the scalar potential are
avoided.  For large values of $\tb$ there is no simple analytical
solution for $A_t$, since the bottom- and tau-Yukawa couplings
cannot be neglected and the fixed point behaviour is  less pronounced.
%
%
%----------------- mu
%
\begin{figure}[th]
  \vspace*{-0.5cm}
  \begin{center}
    \leavevmode
    \epsfxsize=14.7cm
    \epsffile{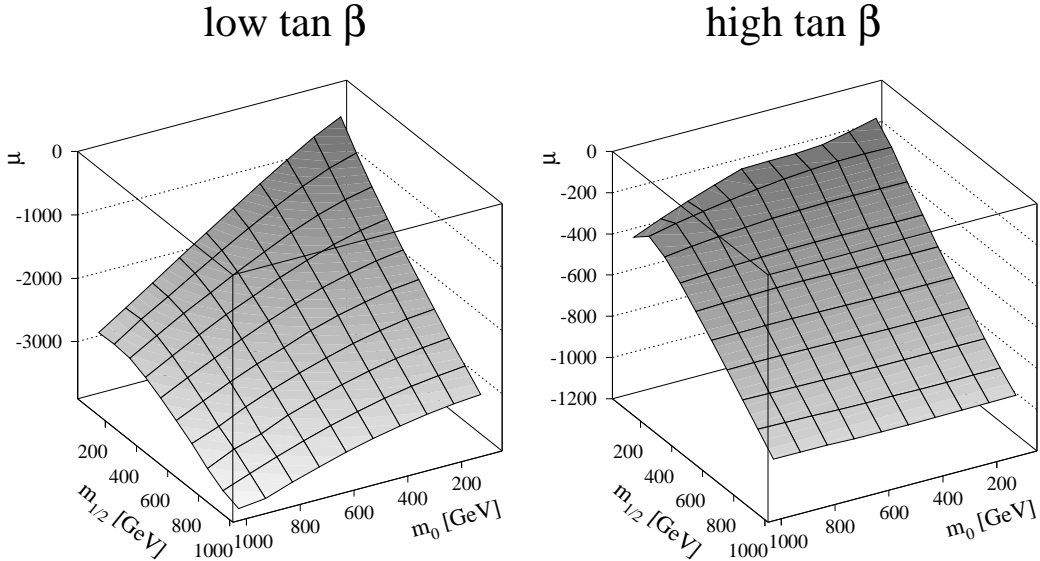}
  \end{center}
  \vspace*{-1.5cm}
  \caption[]{\label{f2}The value of the Higgs mixing parameter
    $\mu$ from EWSB  as function of
    $\mze$ and $\mha$  for the low and high $\tb$ scenario.}
    %RE respectively. }
\end{figure}

The Higgs mixing parameter $\mu$ can be determined from
radiative electroweak symmetry breaking.
Substituting
\beq\label{m1m2}
 m^2_{H_1}&=&m_1^2-\mu^2\\   m^2_{H_2}&=&m_2^2-\mu^2\nonumber
\eeq
into eq.~(\ref{min1}) yields:
\beq
 \mu^2 & = & \frac{m^2_{H_1} - m^2_{H_2}\tan^2\beta}
                  {\tan^2\beta -1} +
             \frac{\Sigma_1  - \Sigma_2 \tan^2\beta}
                  {\tan^2\beta -1}- \frac{M_Z^2}{2}\;\;.
       \label{mufrommz}
\eeq
The dependence of $\mu$ on $m_0$ and $m_{1/2}$ is shown in
fig.~\ref{f2}. Due to the strong $\tb$ dependence in
eq.~(\ref{mufrommz}), the values are much smaller for the high
$\tb$ scenario. One observes that EWSB determines
only $\mu^2$, so the mixing parameter $\tilde{A_t} $ in the
stop sector can be either large or small, depending on the
relative sign of $\mu$ and $A_t$. Both cases will be discussed.

Note that the mass of the CP-odd Higgs boson
%RE rises
increases
%RE
with $\mu^2$
(see eqs.~(\ref{ma}) and (\ref{m1m2})), so the
Higgs mass $m_A$ is in practice always much larger than $M_Z$,
if one determines $\mu$ from EWSB, at least for the low $\tan\beta$
scenario. For large values of $\tan\beta$ $\mu$ can become small,
as can be seen from eq.~(\ref{mufrommz}), in which case
$m_A$ is not necessary small compared with $M_Z$.
Then $m_h$ becomes dependent on $m_A$ and the approximate formula
from eq.~(\ref{higgs2}) breaks down.

%DK new paragraph
%RE One should mention, however, that

Note that the analytical formulae presented here  
 are obtained using the effective potential approach (EPA),
which  samples the Green function at zero momentum and ignores the contribution
to the wave function renormalization (WFR).
The difference between our method and the explicit diagrammatic calculations (FDC )
~\cite{Bri,CPR}  is  at most a few GeV
 as can be seen from the comparison  in tab.~\ref{hlim}.
\begin{table}[ht]
\begin{center}
\begin{tabular}{|c|c|c|c|c|}
\hline
   &                  & Mixing             &
  \multicolumn{2}{|c|}{ $m_h^{max}$ in GeV} \\ \cline{4-5}
                        \rb{$\mze$ in GeV} &
  \rb{$\mha$ in GeV}  & $A_{t,b,\tau}$     &
  EPA                 & FDC \\
\hline
    1000 &   2000 &      0 &  123.5 &  125.8 \\
     200 &   2000 &      0 &   94.2 &   96.9 \\
    1000 &    400 &      0 &  126.0 &  127.6 \\
    1000 &   2000 &      1 &  132.1 &  134.0 \\
\hline
\end{tabular}
\caption[]{\label{hlim} Comparision of EPA Higgs limits with FDC
    calculations \cite{Bri,CPR} for different spectra: all susy
    particles heavy, light squarks/sleptons, light
    charginos/neutralinos, and large mixing in 3.~generation
    sparticles. The mass of the pseudo scalar higgs boson $m_A$ in
    the above examples is $m_A=500$ GeV and the top mass $m_t=180$ GeV.
    For smaller top masses the differences are smaller.
}
\end{center}
\end{table}

\section{Numerical Results}
%
%----------------- ma
%
\begin{figure}[th]
  \vspace*{-1.5cm}
  \begin{center}
    \leavevmode
    \epsfxsize=13.9cm
    \epsffile{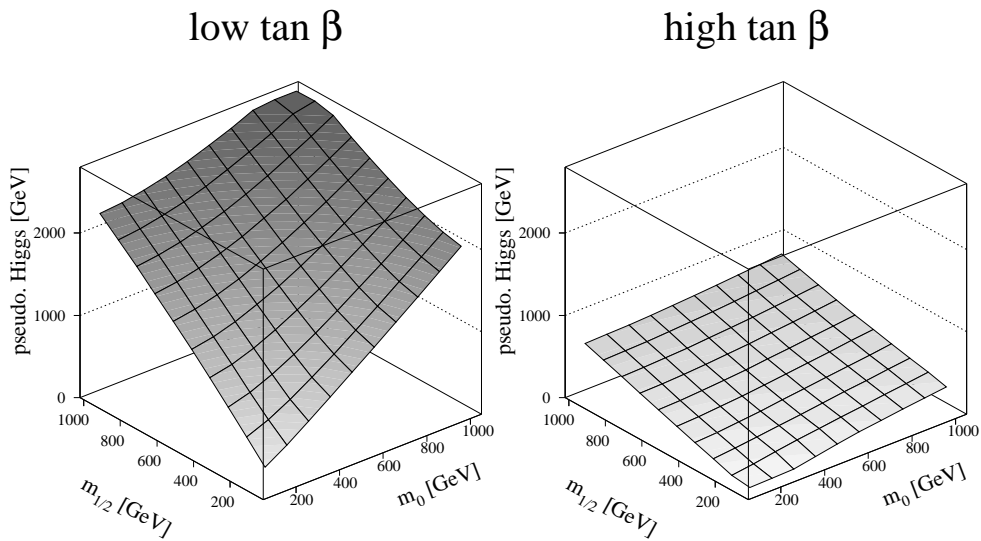}
    \vspace*{-1.8cm}
  \end{center}
  \caption[]{\label{f3}The mass of the CP-odd Higgs as function of
    $\mze$ and $\mha$  for the low and high $\tb$ scenario,
    respectively. The lower values for high $\tb$ originate from
    the lower values of $\mu$ required for EWSB in that
    case~\cite{WE1}.}
\end{figure}

\begin{figure}[th]
  \vspace*{-1.5cm}
  \begin{center}
    \leavevmode
    \epsfxsize=14.0cm
    \epsffile{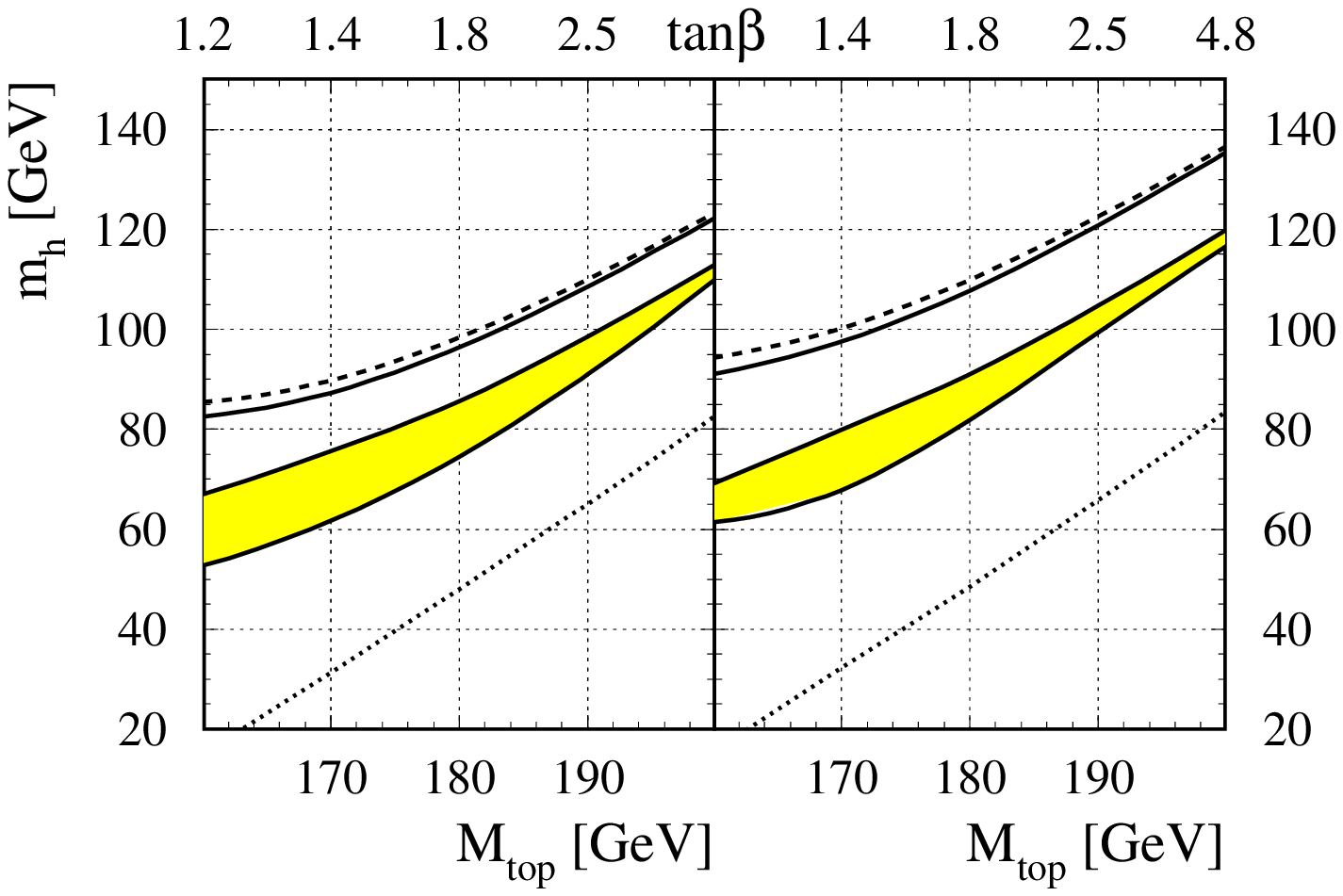}
  \end{center}
  \vspace*{-0.7cm}
  \caption[]{\label{f4}The  mass of the lightest CP-even Higgs as
    function of the top mass at Born level (dotted lines),
    including complete one-loop contributions of all particles
    (dashed lines, $\mu>0$)  for two values of the SUSY masses $\mze$ and $\mha$.
   Two-loop contributions reduce the one-loop
    corrections significantly as shown by the shaded area
    (the upper boundary corresponds to $\mu>0$, the lower one
    to $\mu<0$).  The solid line just
    below the dashed line is the one-loop prediction
    from the third generation only,
    which apparently gives the main contribution. 
    }
\end{figure}

With the Higgs parameters $\mu$ and $\tan\beta$ determined from EWSB and 
bottom-tau unification, one can
calculate all sparticle and Higgs masses as function of $m_0$ and
$m_{1/2}$ including the corrections from all (s)particles in the
loops. Since the CP-odd Higgs mass $m_A$ is
%RE always
for the most part of the parameter space much larger
than $M_Z$ (see fig.~\ref{f3}), the mass of the lightest Higgs
is mainly a function of the top mass only, since the $m_A$ dependence
drops out, as discussed above. 
The top dependence is shown in
fig.~\ref{f4} for various cases. The upper scale indicates
the value of $\tan\beta$ corresponding to the top mass for the low \tb scenario.
In this case the Yukawa couplings $Y_b$ and $Y_\tau$ can be neglected and \tb 
can be directly calculated from the fixed point solution for $Y_t$, which yields
approximately $m_t \approx 205 \sin\beta$.
%
%----------------- mh vs. mtop
%
\begin{table}
\begin{center}
\begin{tabular}{|c|r|r|r|r|r|r|r||r|r|r|r|}
\hline
 \multicolumn{12}{|c|}{Higgs masses in [GeV]}             \\
\hline
&\multicolumn{7}{|c||}{Low $\tb$} & \multicolumn{4}{|c|}{High $\tb$}\\
\hline
& &\multicolumn{3}{|c|}{$\mu < 0$} & \multicolumn{3}{|c||}{$\mu > 0$}
& \multicolumn{4}{|c|}{$\mu < 0$} \\ \hline
Symbol& Born  & $t,b,\tau$ & All & +HO & $t,b,\tau$ & All &
  +HO & Born & $t,b,\tau$ & All & +HO \\
\hline
$m_h$ & 48& 83 & 82 &74 & 98 & 96 & 85& --& 110 & 110 & 107\\
$m_H$ & 649 & 714 & 704 & 715& 706 & 708 & 719& 91& 250 & 230& 250\\
$m_A$ & 644 & 708 & 710 & 707& 702 & 715 & 713& -- & 250 & 230 & 250\\
\hline
$m_h^{max~180}$ & 49& 97 & 95 &82 & 109 & 107 & 91& --& 136 & --  & 117\\
$m_h^{max~190}$ & 67& 114 & 113&99 & 122 & 120 & 105& --& 145 & --  & 122\\
\hline
\end{tabular} \end{center}
 \caption[]{\label{t1}Masses of the
Higgs particles for the small and large $\tan\beta$ scenarios.
The column with $t,b,\tau$ has only contribution from the third
generation of particles, $All$ includes all particles in the loop and $+HO$
includes second order corrections from the renormalization group resummation,
as given in ref. \cite{cw}. The numbers in the first three rows correspond
to the following best fit parameters (in GeV) $m_0=200, m_{1/2}=270,
\mu=\mp 1100, \mu(M_Z)=\mp 525, \tb =1.7 $ 
and $m_0=800, m_{1/2}=90, \mu=-350, \mu(M_Z)=-220, \tb = 41$,
for the low and high $\tb$ scenario,
respectively.
The last rows show  the upper limit   of the lightest
Higgs mass corresponding  to large SUSY masses ($m_0=m_{1/2}$=500 GeV)
for $m_t=180$ and 190 GeV, respectively. The dashes
  for high $\tan\beta$  indicate non-physical negative Higgs masses.
}
\end{table}
%
%----------------- x-section
%
\begin{figure}[ht]
\vspace{-2cm}
\begin{center}
  \leavevmode
  \epsfxsize=14cm
  \epsffile{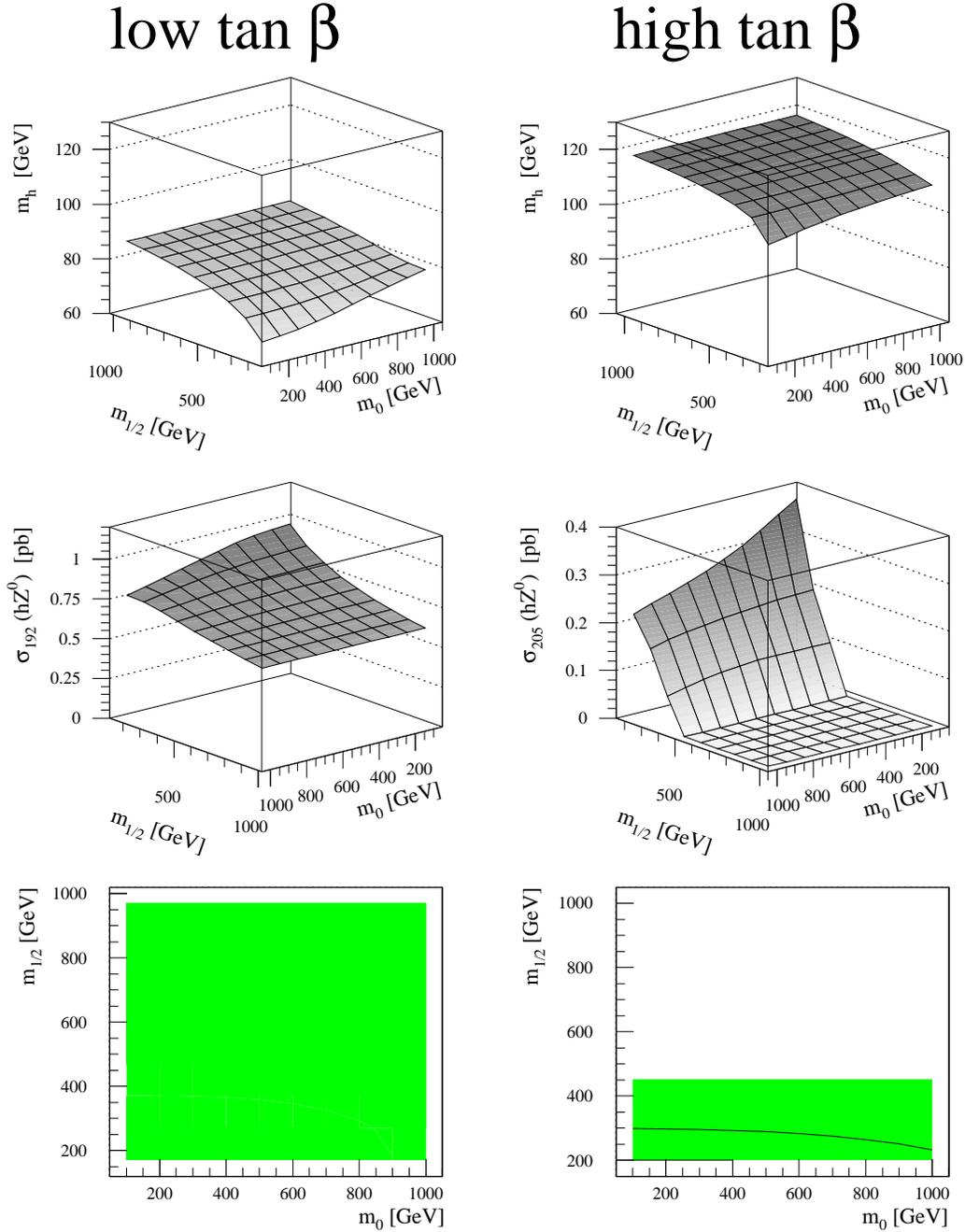}
\end{center}
\caption[]{\label{f5}The mass of the lightest CP-even Higgs as
           function of $\mze$ and $\mha$ for the low and high $\tb$
           scenario on the left and right hand side.
           The top mass was taken to be 179 GeV and
           the sign of $\mu $ is negative,
           as required for the high $\tb$ solution, but chosen
           negative for low $\tb$.
%      A positive value would increase the
%          Higgs mass by 6 to 14 GeV for low \tb, thus lowering
%          the Higgs-strahlungs cross section $e^+e^-\to hZ_0$,
%          shown in the second row, even more.
           For   high $\tan\beta$ the cross section
           is practically zero at the foreseen LEP II energy of 192 GeV,
           so only the cross section at the theoretically  possible
           LEP II energy of 205 GeV is shown. The kinematical reach
           of the parameter space is shown by the contours of the third
           row: the whole parameter space can be covered for
           the low $\tan\beta$ scenario, while for high $\tan\beta$
           the kinematic reach is about $m_{1/2}<450$ GeV and
           the solid line is the contour for $\sigma=0.2$ pb,
           which is roughly the exclusion limit\cite{lepwg}.
                      }
\end{figure}
%
%----------------- x-section vs ecm
%
\begin{figure}[ht]
\vspace{-2cm}
\begin{center}
  \leavevmode
  \epsfxsize=15.8cm
  \epsffile{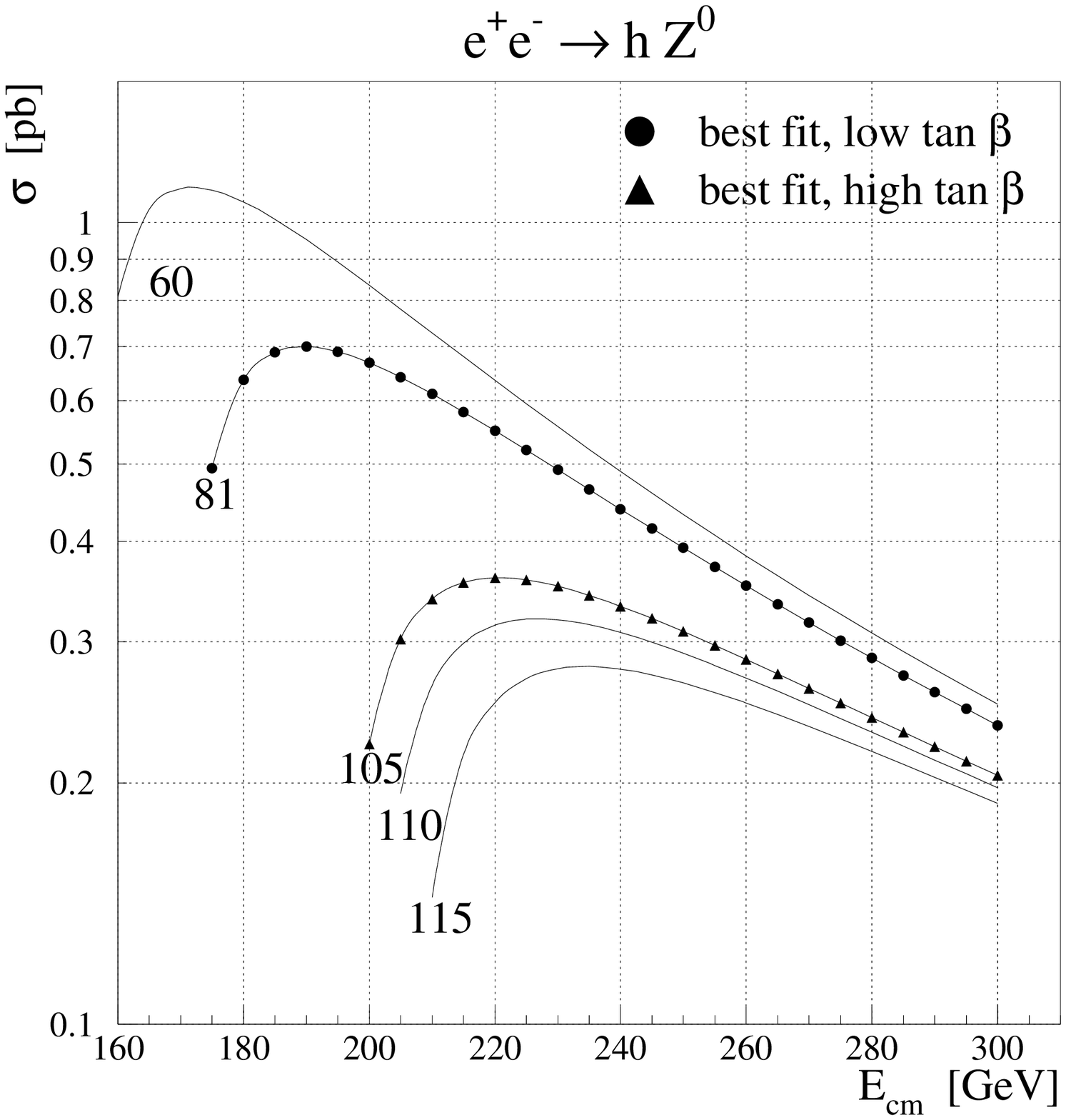}
\end{center}
\caption[]
   {\label{f6}The cross section as function of the center of
    mass energy for different Higgs masses, as indicated by
    the numbers (in GeV). }

\end{figure}

One observes the large one-loop
corrections (upper lines) compared to the Born term (lowest lines)
and the   two-loop corrections\cite{cw} in-between.
The dependence on the sign of $\mu$, which mainly influences
the mixing parameter $\tilde{A}_t$, as discussed above,
is shown by the grey area.
Positive values of $\mu$ are not allowed for the large $\tan\beta$
scenario, because   the large positive corrections to the bottom mass
prohibit bottom-tau unification in that case. These corrections
are small for small values of $\tan\beta$, so in that case one can
choose either sign of $\mu$ (EWSB determines only $\mu^2$). Positive
values of $\mu$ increase   $m_h$ by 6 to 14 GeV, so the sign
is quite important. Note that 
$\mu$ from EWSB is large (see fig. \ref{f2}), so chaning the sign
 corresponds to large changes in the mixing.

 The dominant contributions originate from the Yukawa couplings
of the third generation, as demonstrated by the dashed lines in
fig.~\ref{f4}:
the Higgs masses including the third generation particles in the loops
differ  less than 3 GeV from the masses including the full corrections.
For the large $\tan\beta$ scenario  the largest contribution
originates from the sbottom sector, as shown in  Appendix D, where
the individual contributions from all particles have been summarized.
%RE The masses of all SUSY particles are given too.
The masses of all SUSY particles as obtained from the RG equations
in ref.~\cite{WE1} are given too.

Note the large single contributions to $\Sigma_1$ and $\Sigma_2$.
However, if the masses of the left and right handed partners
are equal, they cancel in the sum of $\Sigma_1$ and $\Sigma_2$.
The contributions from Higgses, charginos, neutralinos and sparticles
of the third generation with a large mixing between left and
right handed partners does not cancel and one obtains large
one-loop corrections.

For the lightest Higgs the large contributions from $\Sigma_1$ and $\Sigma_2$
are partially  cancelled against the corrections in $\Delta_{ij}$, since they
contain similar terms with opposite signs
($\frac{1}{\Psi_{1/2}}\frac{\partial m_k^2}{\partial\Psi_{1/2}}$ in
eqs.~(\ref{Sigma1}), (\ref{Sigma2}) and (\ref{higgs1})).
In the formula for the Higgs mass, eq.~(\ref{higgs}),
these corrections are summed.
Nevertheless, at large $\tb$ the resulting corrections are
still large, as can be seen by adding the bottom lines of tables
\ref{t2}, \ref{t3} and \ref{t4}. Note that the $\Sigma_{1(2)}$
contributions of table \ref{t2}
must be weighted by $\frac{1}{\sin 2\beta} \approx 0.05 $,
as one can see in eqs.~(\ref{min2}) and (\ref{CPevenmas}).

The curves on the left side in fig.~\ref{f4} are shown
for $m_0=200,m_{1/2}=270$, which is the solution yielding
the best fit to the low energy constraints~\cite{WE1}.
On the right hand side $m_0$ and $\mha$ are set to 500 GeV,
which corresponds to squarks masses around 1 TeV.
In this case  one obtains  upper limits on $m_h$,
since  the Higgs mass saturates for such large values, as shown in fig.
\ref{f5}. Here the one-loop Higgs  mass is plotted as function of $\mze$ and $\mha$
for $\mu<0$.  For $\mu>0$ the Higgs mass increases, but the second order
corrections decrease it again by roughly the same  amount, as summarized
numerically in table \ref{t1} for various conditions, so fig. \ref{f5}
is a good representation for the most pessimistic case of $\mu>0$ too,
if the second order corrections are included.

For large $\tan\beta$ the Higgs masses are outside the reach
of LEP II with a maximum energy of 192 GeV. But for the
low $\tan\beta$ scenario the Higgs cross section is large
enough for the whole parameter space of $m_0$ and $m_{1/2}$,
as shown in fig.~\ref{f5}, at least if the top mass is around 180 GeV.
%RE If nature conspires and
%If the top mass is on the high side of the
%experimental results ($M_t>190$ GeV) and the squark masses above 1 TeV,
%then the maximum possible LEP II energy of 205 GeV would be required
%to observe the Higgs boson, even for the low $\tb$ scenario.

The small region of parameter space covered by LEP II at 205 GeV
in case of the high $\tb$ scenario is shown by the solid line
in the bottom part of fig.~\ref{f5}. It is assumed, that a
luminosity of 0.2 pb is needed for the Higgs search, as estimated
by the LEP II working group \cite{lepwg}. At a LEP II energy of
192 GeV the high $\tb$ scenario of the CMSSM is out of reach.
Also the channel $e^+e^-\to hA$ is not accessible at LEP II for
most of the parameter space because of the large values of
$m_A$ predicted in the CMSSM (basically because of the large
value of $\mu$ required by EWSB). The center of mass energy
dependence of the Higgs-strahlungs cross section is summarized
in fig.~\ref{f6} for various Higgs masses.

\section{Summary}
Within the framework of the Minimal Supersymmetric Standard Model
(MSSM) the Higgs masses
and LEP II production cross sections have been calculated including
the D-terms and complete one-loop contributions of all
sparticles, for which explicit analytical formulae are derived.
As expected, the radiative corrections  from the Yukawa
couplings of the third generation dominate over the contributions
from charginos and neutralinos. 
This implies that the simple analytical formula of eq. \ref{higgs2}
is a good approximation. In this case  the maximal Higgs mass is reached
for "maximal" mixing, i.e. maximal $\tilde{X}_t$, which occurs for 
$\tilde{A}_t=\sqrt{6}M_{SUSY}$. In the CMSSM maximal mixing does not
occur, since $\tilde{A}_t\approx M_{SUSY}$, if $A_t$ is determined from 
the fixed point solution for $A_t$ (eq. \ref{aa}) and $\mu$ from 
EWSB (eq. \ref{mufrommz}).

For the low $\tb$ scenario the mass
of the lightest Higgs is found to be below 90 GeV for a top mass
below 180 GeV. The cross section at a LEP II energy
of 192 GeV is sufficient to find or exclude this low $\tb$ scenario.
For the high $\tb$ scenario  only a small fraction of the parameter
space can be covered, since the Higgs mass is predicted
between 105 and 122 GeV in most cases.
It should be noted that the predicted upper limit of  the Higgs mass is 
somewhat lower (up to 10 GeV) than the one  given by the LEP Working Group
\cite{lepwg}, because they assume maximal mixing, which does not occur in the 
CMSSM.
%in the CMSSM  the mixing parameter $\tilde{A}_t$ (eq. \ref{higgs2}) 
%is not maximal.
%, since $A_t$ is fixed by eq. \ref{aa} and $\mu$ by eq. \ref{mufrommz}.

\vspace{0.5cm}

{\em Note added in proof:\\} 
After finishing our results  a new publication containing
all one-loop corrections to all  masses and  couplings 
of the MSSM appeared\cite{Pierce}.
Concerning the Higgs boson masses  our results (within the EPA approach) are more  explicit  and  contain individual contributions coming from different particles together with their  numerical values  for low and high  $\tb$.

 \paragraph{ Acknowledgment}

This work was supported from the Bundesministerium f\"ur Bildung,
Wissenschaft, Forschung und Technologie
(BMBF Contract CERN-LEP-DELPHI 05 6KA16P 3),
the Deutsche Forschungsgemeinschaft (DFG) for our
"Graduiertenkolleg" and the European Community (HCM Network Contract
ERBCHRXCT930345) and Russian Foundation for Basic Research (Grant RFBR-96-02-17379).

\clearpage
\section*{ Appendix A }

In this appendix we summarize the tree-level field-dependent
masses of all particles contributing to the one-loop corrections
to the effective Higgs potential.
\small
\paragraph{Quark and Lepton Masses.}
\begin{eqnarray*}
m_q^2&=&h_q^2|H_i^0|^2, \;\;\;\;\; {\rm where}
\left\{\begin{array}{ll} i=2 & {\rm for}\;\; up\;\; quarks \\
                         i=1 & {\rm for}\;\; down\;\; quarks
\end{array}\right.   \\
m_l^2&=&h_l^2|H_1^0|^2
\end{eqnarray*}

\paragraph{Gauge Boson Masses.}
\begin{eqnarray*}
M^2_{W^{\pm}}&=&\frac{g^2}{2}(|H_1^0|^2+|H_2^0|^2) \\
M^2_{Z^0}&=&\frac{g^2+g'^2}{2}(|H_1^0|^2+|H_2^0|^2)
\end{eqnarray*}

\paragraph{Higgs Boson Masses.}
\begin{eqnarray*} m_A^2&=&m_1^2+m_2^2 \\
m^2_{H^{\pm}}&=&m_A^2+M_W^2 \\
m^2_{H,h}&=&\frac{1}{2}\left[m_A^2+M_Z^2 \pm \sqrt{(m_A^2+M_Z^2)^2
- 4m_A^2M_Z^2\cos^2 2\beta }\right] \\
\end{eqnarray*}

\paragraph{Chargino and Neutralino Masses.}
The chargino mass matrix  is:
\begin{equation}
M^{(c)}=\left(
\begin{array}{cc}
M_2 & g|H_2^0| \\  g|H_1^0|& \mu
\end{array} \right)
\end{equation}
where $M_2$ is the mass of the $SU(2)_L$ gaugino, the Wino.

After diagonalization one obtains:
\begin{eqnarray*}
\tilde m^2_{\chi_{1,2}^{\pm}}&=&\frac{1}{2}\left[M_2^2+
\mu^2+2M_W^2\right. \\
&&\left.\pm \sqrt{(M_2^2-\mu^2)^2+4M_W^4\cos^2 2\beta
+4M_W^2(M_2^2+\mu^2+ 2M_2\mu\sin 2\beta )}\right]
\end{eqnarray*}
The neutralino mass matrix is:
\begin{equation}
M^{(0)}=\left(
\begin{array}{cccc}
M_1 & 0 & -\frac{g'}{\sqrt{2}}|H_1^0| &  \frac{g'}{\sqrt{2}}|H_2^0|\\
0 & M_2 & \frac{g}{\sqrt{2}}|H_1^0| & -\frac{g}{\sqrt{2}}|H_2^0|\\
-\frac{g'}{\sqrt{2}}|H_1^0|& \frac{g}{\sqrt{2}}|H_1^0|  & 0 & -\mu \\
\frac{g'}{\sqrt{2}}|H_2^0| & -\frac{g}{\sqrt{2}}|H_2^0|  & -\mu & 0
\end{array} \right)
\end{equation}
Again $M_2$ is the mass of the Wino, whereas $M_1$ is the mass of the
$U(1)_Y$ gaugino, the Bino.
The neutralino masses $\tilde m_{\chi^0}$ are the roots $\lambda _k$
of quartic equation $F(\lambda )=0$,
where
\begin{eqnarray}
F(\lambda )&=&\lambda ^4-(M_1+M_2)\lambda ^3-(M_Z^2+\mu ^2-M_1M_2)
\lambda ^2  \nonumber \\
&&+[(M_2\sin^2\theta _W+M_1\cos^2\theta _W-\mu\sin 2\beta )
M_Z^2+(M_1+M_2)\mu^2]\lambda  \nonumber \\
&&+[(M_2\sin^2\theta _W+M_1\cos^2\theta _W)\mu M_Z^2
\sin 2\beta -M_1M_2\mu^2]
\end{eqnarray}

\paragraph{Squark and Slepton Masses.}

The top and the bottom squark and tau-slepton mass-squared matrices:
\begin{equation}
  {\cal M}^{\tilde{t}}=\left(
  \begin{array}{cc}
    \tilde{m}_{t_L}^2            & m_t(A_tm_0 - \mu\cot \beta ) \\
    m_t(A_tm_0 - \mu\cot \beta ) & \tilde{m}_{t_R}^2
  \end{array}  \right), \label{Mstop}
\end{equation}
\begin{equation}
  {\cal M}^{\tilde{b}}=\left(
  \begin{array}{cc}
    \tilde{m}_{b_L}^2            & m_b(A_bm_0 - \mu\tan \beta ) \\
    m_b(A_bm_0 - \mu\tan \beta ) & \tilde{m}_{b_R}^2
  \end{array}  \right) , \label{Msbot}
\end{equation}
\begin{equation}
  {\cal M}^{\tilde{\tau}}=\left(
  \begin{array}{cc}
    \tilde{m}_{\tau_L}^2   & m_{\tau}(A_{\tau}m_0 - \mu\tan \beta ) \\
    m_{\tau}(A_{\tau}m_0 - \mu\tan \beta ) & \tilde{m}_{\tau_R}^2
  \end{array}  \right).  \label{Mstau}
\end{equation}

The diagonalization of these matrices yields expressions for the masses
of the third generation squarks and sleptons:
\begin{eqnarray}
  \tilde m_{t_{1,2}}^{2} & = &
     \frac{1}{2}\left(\tilde{m}_{t_L}^2 + \tilde{m}_{t_R}^2\right)
     \mp \sqrt{\frac{1}{4}
         \left(\tilde{m}_{t_L}^2-\tilde{m}_{t_R}^2\right)^2 +
         m_t^2(A_tm_0 - \mu\cot\beta)^2}                      , \\
     \tilde m_{b_{1,2}}^{2} & = &
     \frac{1}{2}\left(\tilde{m}_{b_L}^2 + \tilde{m}_{b_R}^2\right)
     \mp \sqrt{\frac{1}{4}
         \left(\tilde{m}_{b_L}^2-\tilde{m}_{b_R}^2\right)^2 +
         m_b^2(A_bm_0 - \mu\tan\beta)^2}                      , \\
    \tilde m_{\tau_{1,2}}^{2} & = &
      \frac{1}{2}\left(\tilde{m}_{\tau_L}^2+\tilde{m}_{\tau_R}^2\right)
      \mp \sqrt{\frac{1}{4}
         \left(\tilde{m}_{\tau_L}^2-\tilde{m}_{\tau_R}^2\right)^2 +
         m_\tau^2(A_\tau m_0 - \mu\tan\beta)^2}                 .
\end{eqnarray}

The mass eigenstate for the massive third generation sneutrino is:
\begin{equation}
\tilde m_{\nu_{\tau}}^{2}=M_L^2+\frac{1}{2}M_Z^2\cos 2\beta
\end{equation}
where $M_L$ is the breaking mass of the 3.~generation Sleptons of
the $SU(2)_L$ doublet.

\section*{ Appendix B }

We present here analytical formulae for the contributions
from all particles to $\Sigma_1$ and $\Sigma_2$ from
eqs.~(\ref{Sigma1}) and (\ref{Sigma2}). Some of these formulae
can be also  found in ref.~\cite{Many}. In most cases they coincide
with ours, however we have found obvious misprints.

\begin{eqnarray*}
%
%   Standardmodell part
%
\Sigma_1^{W^\pm} & = & \frac{3g^2}{32\pi^2}f(M^2_{W^\pm}) \\
\Sigma_2^{W^\pm} & = & \frac{3g^2}{32\pi^2}f(M^2_{W^\pm} )\\
\Sigma_1^{Z^0}   & = & \frac{3g^2}{64\pi^2\cos^2\theta_W}f(M^2_{Z}) \\
\Sigma_2^{Z^0}   & = & \frac{3g^2}{64\pi^2\cos^2\theta_W}f(M^2_{Z} )\\
%
% charged Higgs part
%
\Sigma_1^{H^\pm} & = & \frac{g^2}{32\pi^2}f(m^2_{H^\pm}) \\
\Sigma_2^{H^\pm} & = & \frac{g^2}{32\pi^2}f(m^2_{H^\pm} )\\
\Sigma_1^{H,h}   & = & \frac{g^2}{128\pi^2\cos^2\theta_W}\left[1 \pm
                       \frac{M_Z^2+m_A^2(1-4\cos 2\beta+2\cos^2 2\beta )}
                       {m_H^2-m_h^2} \right] f(m^2_{H,h})\\
\Sigma_2^{H,h}   & = & \frac{g^2}{128\pi^2\cos^2\theta_W}\left[1 \pm
                       \frac{M_Z^2+m_A^2(1+4\cos 2\beta+2\cos^2 2\beta )}
                       {m_H^2-m_h^2} \right] f(m^2_{H,h})\\
%
%  Chargino part
%
\Sigma_1^{\tilde \chi_{1,2}^{\pm}}
                 & = & -\frac{g^2}{16\pi^2}\left[1\pm
                      \frac{2M_W^2\cos 2\beta +M_2^2+
                      \mu^2+2M_2\mu\tan\beta}
                      {\tilde m^2_{\chi_1^{\pm}}-
                      \tilde m^2_{\chi_2^{\pm}}}\right]
                      f(\tilde m^2_{\chi_{1,2}^{\pm}})\\
\Sigma_2^{\tilde \chi_{1,2}^{\pm}}
                 & = & -\frac{g^2}{16\pi^2}\left[1\pm
                        \frac{-2M_W^2\cos 2\beta +M_2^2+\mu^2+
                        2M_2\mu\cot\beta}{\tilde m^2_{\chi_1^{\pm}}
                       -\tilde m^2_{\chi_2^{\pm}}}\right]
                       f(\tilde m^2_{\chi_{1,2}^{\pm}})\\
%
%   Neutralino
%
\Sigma_1^{\tilde \chi^0}
                 & = & -\frac{g^2}{16\pi^2}\frac{M_Z^2}{M_W^2}
                        \sum\limits_{k=1}^{4}
                        \frac{\lambda_k(\lambda_k-\tilde m_{\gamma})
                        (\lambda_k+\mu\tan\beta)}{D(\lambda_k)}
                       f(\lambda_k^2) \\
\Sigma_2^{\tilde \chi^0}
                 & = & -\frac{g^2}{16\pi^2}\frac{M_Z^2}{M_W^2}
                       \sum\limits_{k=1}^{4}
                       \frac{\lambda_k(\lambda_k-\tilde m_{\gamma})
                       (\lambda_k+\mu\cot\beta)}{D(\lambda_k)}
                       f(\lambda_k^2)\\
%
%  (S)top
%
\Sigma_1^t     & = & 0                                         \\
\Sigma_2^t     & = & -\frac{3g^2}{16\pi\cos^2\theta_W}
                      \frac{m_t^2}{M_Z^2\sin^2\beta}
                      f(m_t^2)                    \\
\Sigma_1^{\tilde t_{1,2}} & = &
  \frac{3g^2}{32\pi^2\cos^2\theta_W}f(\tilde m^2_{t_{1,2}})
  \Biggl\{\frac{1}{4}\pm \frac{1}{\smas{t_1}{2}-\smas{t_2}{2}}\\
& &
%correct misprint 31.1.2001 wdb
%  \times\biggl[\frac{m^2_t \mu}{M^2_Z\sin^2\beta}
  \times\biggl[\frac{-m^2_t \mu}{M^2_Z\sin^2\beta}
  \left(A_tm_0\tan\beta - \mu\right)
  + \frac{1}{2}\left(\frac{1}{2} -\frac{4}{3}\sin^2\theta_W\right)
  \left(\smas{t_L}{2}-\smas{t_R}{2}\right)\biggr]\Biggr\}\nn
\Sigma_2^{\tilde t_{1,2}} & = &
  \frac{3g^2}{32\pi^2\cos^2\theta_W}f(\tilde m^2_{t_{1,2}})
  \Biggl\{\frac{m^2_t}{M^2_Z\sin^2\beta}-\frac{1}{4}\pm
    \frac{1}{\smas{t_1}{2}-\smas{t_2}{2}}\\
& &
  \times\biggl[\frac{m^2_tA_tm_0}{M^2_Z\sin^2\beta}
  \left(A_tm_0-\mu\cot\beta\right)-\frac{1}{2}
  \left(\frac{1}{2}-\frac{4}{3}\sin^2\theta_W\right)
  \left(\smas{t_L}{2}-\smas{t_R}{2}\right)\biggr]\Biggr\}\nn
%com\eeq
%
%     bottom Beitraege
%
%com\beq
\Sigma_1^b     & = & -\frac{3g^2}{16\pi\cos^2\theta_W}
                      \frac{m_b^2}{M_Z^2\cos^2\beta}
                      f(m_b^2)                    \\
\Sigma_2^b     & = & 0                                         \\
\Sigma_1^{\tilde b_{1,2}}&=&\frac{3g^2}{32\pi^2\cos^2\theta_W}
  f(\tilde m^2_{b_{1,2}})
  \Biggl\{\frac{m^2_b}{M^2_Z\cos^2\beta} -\frac{1}{4}
  \pm\frac{1}{\smas{b_1}{2}-\smas{b_2}{2}}\\
& &
  \times\biggl[\frac{m^2_bA_bm_0}{M^2_Z\cos^2\beta}
  \left(A_bm_0-\mu\tan\beta\right)-\frac{1}{2}\left(\frac{1}{2}-\frac{2}{3}
  \sin^2\theta_W\right)\left(\smas{b_L}{2}-
  \smas{b_R}{2}\right)\biggr]\Biggr\}\nn
\Sigma_2^{\tilde b_{1,2}}&=&\frac{3g^2}{32\pi^2\cos^2\theta_W}
  f(\tilde m^2_{b_{1,2}})
  \Biggl\{\frac{1}{4}\pm\frac{1}{\smas{b_1}{2}-\smas{b_2}{2}}\\
 & &
  \times\biggl[\frac{-m^2_b\mu}{M^2_Z\cos^2\beta}
  \left(A_bm_0\cot\beta-\mu\right)+\frac{1}{2}\left(\frac{1}{2}-\frac{2}{3}
  \sin^2\theta_W\right)\left(\smas{b_L}{2}-
  \smas{b_R}{2}\right)\biggr]\Biggr\}\nn
%com\eeq
%
%     tau Beitraege
%
%com\beq
\Sigma_1^\tau  & = & -\frac{g^2}{16\pi\cos^2\theta_W}
                      \frac{m_{\tau}^2}{M_Z^2\cos^2\beta}
                      f(m_{\tau}^2)                            \\
\Sigma_2^\tau  & = & 0                                         \\
\Sigma_1^{\tilde \tau_{1,2}}&=&\frac{3g^2}{32\pi^2\cos^2\theta_W}
  f(\tilde m^2_{\tau_{1,2}})
  \Biggl\{\frac{m^2_{\tau}}{M^2_Z\cos^2\beta} -\frac{1}{4}
  \pm\frac{1}{\smas{\tau_1}{2}-\smas{\tau_2}{2}}\\
& &
  \times\biggl[\frac{m^2_{\tau} A_{\tau} m_0}{M^2_Z\cos^2\beta}
  \left(A_{\tau} m_0-\mu\tan\beta\right)-\frac{1}{2}\left(\frac{1}{2}-
  2\sin^2\theta_W\right)\left(\smas{\tau_L}{2}-
  \smas{\tau_R}{2}\right)\biggr]\Biggr\}\nn
\Sigma_2^{\tilde \tau_{1,2}}&=&\frac{3g^2}{32\pi^2\cos^2\theta_W}
  f(\tilde m^2_{\tau_{1,2}})
  \Biggl\{\frac{1}{4}\pm\frac{1}{\smas{\tau_1}{2}-\smas{\tau_2}{2}}\\
 & &
  \times\biggl[\frac{-m^2_{\tau}\mu}{M^2_Z\cos^2\beta}
  \left(A_{\tau}m_0\cot\beta-\mu\right)+
  \frac{1}{2}\left(\frac{1}{2}-\frac{2}{3}
  \sin^2\theta_W\right)\left(\smas{\tau_L}{2}-
  \smas{\tau_R}{2}\right)\biggr]\Biggr\}\nn
\end{eqnarray*}
%\vglue 1cm
%
%\clearpage
\section*{ Appendix C }
%\vspace*{-0.5cm}
Here analytical formulae for the contributions from all particles to
$\Delta_{11}, \Delta_{22}$ and $\Delta_{12}$ in eq.~(\ref{higgs1})
are presented.
%RE
In the literature often $m_A$ is considered as a free parameter.
In our framework this can be achieved, if the CP-even mass matrix
eq.~(\ref{Mmh}) is rewritten as:
\beq
{\cal M}=\frac{1}{2}
\left(\frac{\partial^2 V}{\partial \psi_i \partial \psi_j}
\right)_{v_1,v_2}&=&
\frac{1}{2}\left(\begin{array}{cc}
\tan\beta & -1 \\
-1 & \cot\beta
\end{array}\right) m_{A_{1-loop}}^2\sin 2\beta +\frac{1}{2}
\left(\begin{array}{cc}
\cot\beta & -1 \\
-1 & \tan\beta
\end{array}\right) M_Z^2\sin 2\beta   \nn
&+& \frac{1}{2}
\left(\begin{array}{cc}
\Delta_{11}-\tan\beta\;\Delta & \Delta_{12}+\Delta      \\
\Delta_{12}+\Delta            & \Delta_{22}-\cot\beta\;\Delta
\end{array}\right)
\eeq
where $\Delta$ are the corrections from eq.~(\ref{macorr}) and
$\sin 2\beta \;m_{A_{1-loop}}^2=2m_3^2 + \Delta $ (\ref{ma}).
Thus taking $m_A$ as free input parameter, one has to subtract the
terms proportional $\Delta$ in $\Delta_{11}$, $\Delta_{12}$,
and $\Delta_{22}$, having in mind, that they are
compensated with the corresponding corrections in $m_A$.

\paragraph{Contribution from $W^{\pm},Z^0$.}

\begin{eqnarray*}
\Delta_{11}&=&\frac{3g^2}{16\pi^2}M_Z^2
\left[(2\cos^2\theta_W+\cos^{-2}\theta_W)\log\frac{M_Z^2}{Q^2} +
2\cos^2\theta_W\log\cos^2\theta_W \right]\cos^2\beta  \\
\Delta_{22}&=&\frac{3g^2}{16\pi^2}M_Z^2
\left[(2\cos^2\theta_W+\cos^{-2}\theta_W)\log\frac{M_Z^2}{Q^2} +
2\cos^2\theta_W\log\cos^2\theta_W \right]\sin^2\beta  \\
\Delta_{12}&=&\frac{3g^2}{16\pi^2}M_Z^2
\left[(2\cos^2\theta_W+\cos^{-2}\theta_W)\log\frac{M_Z^2}{Q^2} +
2\cos^2\theta_W\log\cos^2\theta_W \right]\sin\beta\cos\beta  \\
\end{eqnarray*}

\paragraph{Contribution from $H^{\pm}$.}

\begin{eqnarray*}
\Delta_{11}&=&\frac{g^2}{8\pi^2}M_W^2
\log\frac{m_A^2+M_W^2}{Q^2}\cos^2\beta  \\
\Delta_{22}&=&\frac{g^2}{8\pi^2}M_W^2
\log\frac{m_A^2+M_W^2}{Q^2}\sin^2\beta  \\
\Delta_{12}&=&\frac{g^2}{8\pi^2}M_W^2
\log\frac{m_A^2+M_W^2}{Q^2}\sin\beta\cos\beta
\end{eqnarray*}

\paragraph{Contribution from H,h.}

\begin{eqnarray*}
\Delta _{11}&=&\frac{g^2}{64\pi^2}\frac{M_Z^2}{M_W^2}\left[ M_Z^2
\left( \log \frac{m_H^2 m_h^2} {Q^4} + 2\rho_1 \log \frac{m_H^2}{m_h^2}
+ \rho_1^2 d(m_H^2,m_h^2)\right) \right.\\
&&  + \left.2(M_Z^2-16m_A^2 \sin^4\beta)\frac{f(m_H^2)
-f(m_h^2)}{m_H^2-m_h^2}\right]\cos^2\beta \\
\Delta _{22}&=&\frac{g^2}{64\pi^2}\frac{M_Z^2}{M_W^2} \left[ M_Z^2
\left( \log \frac{m_H^2 m_h^2}{Q^4} + 2\rho_2 \log \frac{m_H^2}{m_h^2}
+ \rho_2^2 d(m_H^2,m_h^2) \right) \right.   \\
& & + \left. 2(M_Z^2-16m_A^2 \cos^4 \beta)
\frac{f(m_H^2)-f(m_h^2)}{m_H^2-m_h^2}\right]\sin^2 \beta \\
\Delta _{12}&=&\frac{g^2}{64\pi^2}\frac{M_Z^2}{M_W^2} \left[ M_Z^2
\left( \log \frac{m_H^2 m_h^2}{Q^4} + (\rho_1+\rho_2) \log
\frac{m_H^2}{m_h^2} + \rho_1\rho_2 d(m_H^2,m_h^2)\right) \right.   \\
& & + \left. 2(M_Z^2+16m_A^2 \sin^2 \cos^2 \beta)
\frac{f(m_H^2)-f(m_h^2)}{m_H^2-m_h^2}\right]\sin\beta \cos\beta
\end{eqnarray*}
where
\begin{eqnarray*}
\rho_1&=&\frac{M_Z^2+m_A^2(1-4\cos 2\beta + 2\cos^2 2\beta )}
{m_H^2 - m_h^2} \\
\rho_2&=&\frac{M_Z^2+m_A^2(1+4\cos 2\beta + 2\cos^2 2\beta )}
{m_H^2 - m_h^2}
\end{eqnarray*}
$$ d(m_1^2,m_2^2)=2-\frac{m_1^2+m_2^2}{m_1^2-m_2^2}
\log\frac{m_1^2}{m_2^2} $$

\paragraph{Chargino contribution.}

\begin{eqnarray*}
\Delta _{11}&=&-\frac{g^2}{4\pi^2}\left[ M_W^2 \left( \log
\frac{m_{\chi_{1}^{\pm}}^2 m_{\chi_{2}^{\pm}}^2} {Q^4} + 2\eta_1 \log
\frac{m_{\chi_{1}^{\pm}}^2}{m_{\chi_{2}^{\pm}}^2} + \eta_1^2
d(m_{\chi_{1}^{\pm}}^2,m_{\chi_{2}^{\pm}}^2)\right) \right.\\
&&  + \left. \left(2M_W^2-M_2\mu\frac{\sin\beta }{\cos^3\beta}\right)
\frac{f(m_{\chi_{1}^{\pm}}^2)-f(m_{\chi_{2}^{\pm}}^2)}
{m_{\chi_{1}^{\pm}}^2-m_{\chi_{2}^{\pm}}^2}\right]\cos^2\beta \\
\Delta _{22}&=&-\frac{g^2}{4\pi^2}\left[ M_W^2 \left( \log
\frac{m_{\chi_{1}^{\pm}}^2 m_{\chi_{2}^{\pm}}^2} {Q^4} + 2\eta_2 \log
\frac{m_{\chi_{1}^{\pm}}^2}{m_{\chi_{2}^{\pm}}^2} + \eta_2^2
d(m_{\chi_{1}^{\pm}}^2,m_{\chi_{2}^{\pm}}^2)\right) \right.\\
&&  + \left.\left(2M_W^2-M_2\mu\frac{\cos\beta }{\sin^3\beta}\right)
\frac{f(m_{\chi_{1}^{\pm}}^2)-f(m_{\chi_{2}^{\pm}}^2)}
{m_{\chi_{1}^{\pm}}^2-m_{\chi_{2}^{\pm}}^2}\right]\sin^2\beta \\
\Delta _{12}&=&-\frac{g^2}{4\pi^2}\left[ M_W^2 \left( \log
\frac{m_{\chi_{1}^{\pm}}^2 m_{\chi_{2}^{\pm}}^2} {Q^4} +
(\eta_1+\eta_2) \log \frac{m_{\chi_{1}^{\pm}}^2}{m_{\chi_{2}^{\pm}}^2}
+ \eta_1\eta_2 d(m_{\chi_{1}^{\pm}}^2,m_{\chi_{2}^{\pm}}^2)\right)
\right.\\ &&  -
\left.\left(2M_W^2-M_2\mu\frac{1}{\sin\beta\cos\beta}\right)
\frac{f(m_{\chi_{1}^{\pm}}^2)-f(m_{\chi_{2}^{\pm}}^2)}
{m_{\chi_{1}^{\pm}}^2-m_{\chi_{2}^{\pm}}^2}\right]\sin\beta\cos\beta
\end{eqnarray*}
where
\begin{eqnarray*}
\eta_1&=&\frac{2M_W^2\cos 2\beta+M_2^2+\mu^2+2M_2\mu\tan\beta}
{m_{\chi_{1}^{\pm}}^2-m_{\chi_{2}^{\pm}}^2}  \\
\eta_2&=&\frac{-2M_W^2\cos 2\beta+M_2^2+\mu^2+2M_2\mu\cot\beta}
{m_{\chi_{1}^{\pm}}^2-m_{\chi_{2}^{\pm}}^2}
\end{eqnarray*}

\paragraph{Neutralino contribution.}

\begin{eqnarray*}
\Delta_{11}&=&-\frac{g^2}{8\pi^2}\frac{M_Z^2}{M_W^2}
\cos^2\beta \sum_{k=1}^{4}\left[2M_Z^2\left(
    2\lambda _k^2\frac{\tau _1^2(\lambda_k)}{D^2(\lambda_k)}
    \log\frac{\lambda _k^2}{Q^2}+\frac{\tau _1^2
      \left(1-\lambda _k\frac{D'(\lambda _k)}{D(\lambda _k)}\right)}
    {D^2(\lambda_k)}f(\lambda_k^2)
  \right. \right.   \\
&& \left. \left.
+\frac{
2\tau _1\lambda _k(2\lambda _k-\tilde m_{\gamma}+
\mu \tan\beta )}{D^2(\lambda _k)}
f(\lambda _k^2) \right)
-\frac{\lambda _k\mu (\lambda _k-
\tilde m_{\gamma})}{D(\lambda _k)}
\frac{\sin\beta }{\cos^3\beta }
f(\lambda _k^2) \right]  \\
\Delta_{22}&=&-\frac{g^2}{8\pi^2}\frac{M_Z^2}{M_W^2}
\sin^2\beta \sum_{k=1}^{4}\left[2M_Z^2\left(
2\lambda _k^2\frac{\tau _2^2(\lambda_k)}{D^2(\lambda_k)}
\log\frac{\lambda _k^2}{Q^2}
+\frac{\tau _2^2\left(1-\lambda _k\frac{D'(\lambda _k)}
    {D(\lambda _k)}\right)}
{D^2(\lambda_k)}f(\lambda_k^2)
\right. \right.   \\
&& \left. \left.
+\frac{
2\tau _2\lambda _k(2\lambda _k-\tilde m_{\gamma}-\mu \cot\beta )}
{D^2(\lambda _k)}
f(\lambda _k^2) \right)
-\frac{\lambda _k\mu (\lambda _k-
\tilde m_{\gamma})}{D(\lambda _k)}\frac{\cos\beta }{\sin^3\beta }
f(\lambda _k^2) \right]   \\
\Delta_{12}&=&-\frac{g^2}{8\pi^2}\frac{M_Z^2}{M_W^2}
\sin\beta \cos\beta
\sum_{k=1}^{4}\left[2M_Z^2\left( 2\lambda
_k^2\frac{\tau_1\tau_2}{D^2}\log\frac{\lambda _k^2}{Q^2}
+\frac{\tau _1\tau_2\left(1-\lambda
_k\frac{D'}{D}\right)}
{D^2(\lambda_k)}f(\lambda_k^2)
\right.\right.   \\
&&  \left.
+\frac{
\lambda _k[\tau _1(2\lambda _k-\tilde m_{\gamma}+\mu \cot\beta )
+\tau_2(2\lambda_k-\tilde m_{\gamma}+\mu\tan\beta)]}{D^2(\lambda _k)}
f(\lambda _k^2) \right)  \\
&& \left. +\frac{\lambda _k\mu
(\lambda _k-\tilde m_{\gamma})}{D(\lambda
_k)}\frac{1}{\sin\beta \cos\beta } f(\lambda _k^2) \right]
\end{eqnarray*}
where
\begin{eqnarray*}
\tau _1(\lambda _k)&=&
(\lambda _k-\tilde m_{\gamma})
(\lambda_k+\mu \tan \beta) \\
\tau _2(\lambda _k)&=&
(\lambda _k-\tilde m_{\gamma})
(\lambda_k+\mu \cot \beta) \\
\tilde m_{\gamma}&=&
M_2\sin^2\theta _W+M_1\cos^2\theta _W \\
D(\lambda )&=&\frac{\partial F(\lambda )}{\partial \lambda } \ \
D'(\lambda )=\frac{\partial D(\lambda )}{\partial \lambda }
\end{eqnarray*}

\paragraph{Top-stop contribution.}

\begin{eqnarray*}
\Delta_{11}&=&\frac{3g^2}{32\pi^2M_W^2\cos^2\beta}\left[\left(
\frac{m_Z^2}{2}\cos^2\beta\right)^2
\log\frac{\tilde m_{t_1}^2\tilde m_{t_2}^2}{Q^4} \right. \\
&&+2\left(\frac{m_Z^2}{2}\cos^2\beta\right)\xi_1
\log\frac{\tilde m_{t_1}^2}{\tilde m_{t_2}^2}+
\xi_1^2d(\tilde m_{t_1}^2,\tilde m_{t_2}^2)  \\
&&+\left.\left(\frac{1}{18}(8M_W^2-5M_Z^2)^2\cos^4\beta
+2m_t^2A_tm_0\mu\cot\beta \right)
\frac{f(\tilde m_{t_1}^2)-f(\tilde m_{t_2}^2)}
{\tilde m_{t_1}^2-\tilde m_{t_2}^2}\right]   \\
\Delta_{22}&=&\frac{3g^2}{32\pi^2M_W^2\sin^2\beta}\left[
-8m_t^4\log\frac{m_t^2}{Q^2}+
\left(2m_t^2-\frac{m_Z^2}{2}
\sin^2\beta\right)^2\log\frac{\tilde m_{t_1}^2\tilde m_{t_2}^2}{Q^4}
\right. \\
&&+2\left(2m_t^2-\frac{m_Z^2}{2}\sin^2\beta\right)\xi_2
\log\frac{\tilde m_{t_1}^2}{\tilde m_{t_2}^2}+
\xi_2^2d(\tilde m_{t_1}^2,\tilde m_{t_2}^2) \\
&&+\left.\left(\frac{1}{18}(8M_W^2-5M_Z^2)^2\sin^4\beta
+2m_t^2A_tm_0\mu\cot\beta \right)
\frac{f(\tilde m_{t_1}^2)-f(\tilde m_{t_2}^2)}
{\tilde m_{t_1}^2-\tilde m_{t_2}^2}\right]   \\
\Delta_{12}&=&\frac{3g^2}{32\pi^2M_W^2\sin\beta\cos\beta}\left[
\frac{M_Z^2}{2}\cos^2\beta \left(2m_t^2-\frac{m_Z^2}{2}
\sin^2\beta\right) \log\frac{\tilde m_{t_1}^2\tilde m_{t_2}^2}{Q^4}
\right. \\
&&+\left(\frac{M_Z^2}{2}\cos^2\beta\xi_2+
\left(2m_t^2-\frac{m_Z^2}{2}\sin^2\beta\right)\xi_1\right)
\log\frac{\tilde m_{t_1}^2}{\tilde m_{t_2}^2}+
\xi_1\xi_2d(\tilde m_{t_1}^2,\tilde m_{t_2}^2) \\
&&+\left.\left(-\frac{1}{18}(8M_W^2-5M_Z^2)^2\sin^2\beta\cos^2\beta
-2m_t^2A_tm_0\mu\cot\beta \right)
\frac{f(\tilde m_{t_1}^2)-f(\tilde m_{t_2}^2)}
{\tilde m_{t_1}^2-\tilde m_{t_2}^2}\right]
\end{eqnarray*}
where
\begin{eqnarray*}
\xi_1&=&\frac{1}{\tilde m_{t_1}^2-\tilde m_{t_2}^2}
\bigg[-2m_t^2\mu\cot\beta(A_tm_0-\mu\cot\beta)
 +\frac{1}{6}(8M_W^2-5M_Z^2) \left(\smas{t_L}{2}-\smas{t_R}{2}\right)
\cos^2\beta \bigg] \\
\xi_2&=&\frac{1}{\tilde m_{t_1}^2-\tilde m_{t_2}^2}
\bigg[2m_t^2A_tm_0(A_tm_0-\mu\cot\beta)
 -\frac{1}{6}(8M_W^2-5M_Z^2) \left(\smas{t_L}{2}-\smas{t_R}{2}\right)
\sin^2\beta \bigg]
\end{eqnarray*}

\paragraph{Bottom-sbottom contribution.}

\begin{eqnarray*}
\Delta_{11}&=&\frac{3g^2}{32\pi^2M_W^2\cos^2\beta}\left[
-8m_b^4\log\frac{m_b^2}{Q^2}+
\left(2m_b^2-\frac{m_Z^2}{2}
\cos^2\beta\right)^2\log\frac{\tilde m_{b_1}^2\tilde m_{b_2}^2}{Q^4}
\right. \\
&&+2\left(2m_b^2-\frac{m_Z^2}{2}\cos^2\beta\right)\kappa_1
\log\frac{\tilde m_{b_1}^2}{\tilde m_{b_2}^2}+
\kappa_1^2d(\tilde m_{b_1}^2,\tilde m_{b_2}^2) \\
&&+\left.\left(\frac{1}{18}(4M_W^2-M_Z^2)^2\cos^4\beta
+2m_b^2A_bm_0\mu\tan\beta \right)
\frac{f(\tilde m_{b_1}^2)-f(\tilde m_{b_2}^2)}
{\tilde m_{b_1}^2-\tilde m_{b_2}^2}\right]   \\
\Delta_{22}&=&\frac{3g^2}{32\pi^2M_W^2\sin^2\beta}\left[
\left(\frac{m_Z^2}{2}\sin^2\beta\right)^2\log\frac{\tilde
m_{b_1}^2\tilde m_{b_2}^2}{Q^4}  \right. \\
&& +2\left(\frac{m_Z^2}{2}\sin^2\beta\right)\kappa_2
\log\frac{\tilde m_{b_1}^2}{\tilde m_{b_2}^2}+
\kappa_2^2d(\tilde m_{b_1}^2,\tilde m_{b_2}^2) \\
&&+\left.\left(\frac{1}{18}(4M_W^2-M_Z^2)^2\sin^4\beta
+2m_b^2A_bm_0\mu\tan\beta \right)
\frac{f(\tilde m_{b_1}^2)-f(\tilde m_{b_2}^2)}
{\tilde m_{b_1}^2-\tilde m_{b_2}^2}\right]   \\
\Delta_{12}&=&\frac{3g^2}{32\pi^2M_W^2\sin\beta\cos\beta}\left[
\frac{M_Z^2}{2}\sin^2\beta \left(2m_b^2-\frac{m_Z^2}{2}
\cos^2\beta\right) \log\frac{\tilde m_{b_1}^2\tilde m_{b_2}^2}{Q^4}
\right. \\
&&+\left(\frac{M_Z^2}{2}\sin^2\beta\kappa_1+
\left(2m_b^2-\frac{m_Z^2}{2}\cos^2\beta\right)\kappa_2\right)
\log\frac{\tilde m_{b_1}^2}{\tilde m_{b_2}^2}+
\kappa_1\kappa_2 d(\tilde m_{b_1}^2,\tilde m_{b_2}^2) \\
&&+\left.\left(-\frac{1}{18}(4M_W^2-M_Z^2)^2\sin^2\beta\cos^2\beta
-2m_b^2A_bm_0\mu\tan\beta \right)
\frac{f(\tilde m_{b_1}^2)-f(\tilde m_{b_2}^2)}
{\tilde m_{b_1}^2-\tilde m_{b_2}^2}\right]
\end{eqnarray*}
where
\begin{eqnarray*}
\kappa_1&=&\frac{1}{\tilde m_{b_1}^2-\tilde m_{b_2}^2}
\bigg[2m_b^2A_bm_0(A_bm_0-\mu\tan\beta)
  -\frac{1}{6}(4M_W^2-M_Z^2)\left(\smas{b_L}{2}-\smas{b_R}{2}\right)
\cos^2\beta \bigg] \\
\kappa_2&=&\frac{1}{\tilde m_{b_1}^2-\tilde m_{b_2}^2}
\bigg[-2m_b^2\mu\tan\beta(A_bm_0-\mu\tan\beta)
 +\frac{1}{6}(4M_W^2-M_Z^2)\left(\smas{b_L}{2}-\smas{b_R}{2}\right)
\sin^2\beta \bigg]
\end{eqnarray*}

\paragraph{Tau lepton-slepton contribution.}

\begin{eqnarray*}
\Delta_{11}&=&\frac{g^2}{32\pi^2M_W^2\cos^2\beta}\left[
-8m_{\tau}^4\log\frac{m_{\tau}^2}{Q^2}+
\left(2m_{\tau}^2-\frac{m_Z^2}{2}
\cos^2\beta\right)^2\log\frac{\tilde m_{\tau_1}^2\tilde
m_{\tau_2}^2}{Q^4} \right. \\
&&+2\left(2m_{\tau}^2-\frac{m_Z^2}{2}\cos^2\beta\right)\omega_1
\log\frac{\tilde m_{\tau_1}^2}{\tilde m_{\tau_2}^2}+
\omega_1^2d(\tilde m_{\tau_1}^2,\tilde m_{\tau_2}^2) \\
&&+\left.\left(\frac{1}{2}(4M_W^2-3M_Z^2)^2\cos^4\beta
+2m_{\tau}^2A_{\tau}m_0\mu\tan\beta \right)
\frac{f(\tilde m_{\tau_1}^2)-f(\tilde m_{\tau_2}^2)}
{\tilde m_{\tau_1}^2-\tilde m_{\tau_2}^2}\right]   \\
\Delta_{22}&=&\frac{g^2}{32\pi^2M_W^2\sin^2\beta}\left[
\left(\frac{m_Z^2}{2}\sin^2\beta\right)^2\log\frac{\tilde
m_{\tau_1}^2\tilde m_{\tau_2}^2}{Q^4}  \right. \\
&& +2\left(\frac{m_Z^2}{2}\sin^2\beta\right)\omega_2
\log\frac{\tilde m_{\tau_1}^2}{\tilde m_{\tau_2}^2}+
\omega_2^2d(\tilde m_{\tau_1}^2,\tilde m_{\tau_2}^2) \\
&&+\left.\left(\frac{1}{2}(4M_W^2-3M_Z^2)^2\sin^4\beta
+2m_{\tau}^2A_{\tau}m_0\mu\tan\beta \right)
\frac{f(\tilde m_{\tau_1}^2)-f(\tilde m_{\tau_2}^2)}
{\tilde m_{\tau_1}^2-\tilde m_{\tau_2}^2}\right]   \\
\Delta_{12}&=&\frac{g^2}{32\pi^2M_W^2\sin\beta\cos\beta}\left[
\frac{M_Z^2}{2}\sin^2\beta \left(2m_{\tau}^2-\frac{m_Z^2}{2}
\cos^2\beta\right) \log\frac{\tilde m_{\tau_1}^2\tilde m_{\tau_2}^2}{Q^4}
\right. \\
&&+\left(\frac{M_Z^2}{2}\sin^2\beta\omega_1+
\left(2m_{\tau}^2-\frac{m_Z^2}{2}\cos^2\beta\right)\omega_2\right)
\log\frac{\tilde m_{\tau_1}^2}{\tilde m_{\tau_2}^2}+
\omega_1\omega_2 d(\tilde m_{\tau_1}^2,\tilde m_{\tau_2}^2) \\
&&+\left.\left(-\frac{1}{2}(4M_W^2-3M_Z^2)^2\sin^2\beta\cos^2\beta
-2m_{\tau}^2A_{\tau}m_0\mu\tan\beta \right)
\frac{f(\tilde m_{\tau_1}^2)-f(\tilde m_{\tau_2}^2)}
{\tilde m_{\tau_1}^2-\tilde m_{\tau_2}^2}\right]
\end{eqnarray*}
where
\begin{eqnarray*}
\omega_1&=&\frac{1}{\tilde m_{\tau_1}^2-\tilde m_{\tau_2}^2}
\bigg[2m_{\tau}^2A_{\tau}m_0(A_{\tau}m_0-\mu\tan\beta)
 -\frac{1}{2}(4M_W^2-3M_Z^2)
\left(\smas{\tau_L}{2}-\smas{\tau_R}{2}\right)
\cos^2\beta \bigg] \\
\omega_2&=&\frac{1}{\tilde m_{\tau_1}^2-\tilde m_{\tau_2}^2}
\bigg[-2m_{\tau}^2\mu\tan\beta(A_{\tau}m_0-\mu\tan\beta)
 +\frac{1}{2}(4M_W^2-3M_Z^2)
\left(\smas{\tau_L}{2}-\smas{\tau_R}{2}\right)
\sin^2\beta \bigg]
\end{eqnarray*}

\paragraph{Tau-sneutrino contribution.}

\begin{eqnarray*}
\Delta_{11}&=&\frac{g^2}{32\pi^2}\frac{M_Z^4}{M_W^2}
\cos^2\beta \log\frac{\tilde m_{\nu_{\tau}}^2}{Q^2}  \\
\Delta_{22}&=&\frac{g^2}{32\pi^2}\frac{M_Z^4}{M_W^2}
\sin^2\beta \log\frac{\tilde m_{\nu_{\tau}}^2}{Q^2}  \\
\Delta_{12}&=&-\frac{g^2}{32\pi^2}\frac{M_Z^4}{M_W^2}
\sin\beta\cos\beta \log\frac{\tilde m_{\nu_{\tau}}^2}{Q^2}
\end{eqnarray*}

\paragraph{First and second generations.}

Contributions originating from the particles of the two light
generations can be easily reproduced replacing $t \rightarrow u,s$;
$b \rightarrow d,c$; $\tau \rightarrow e,\mu $, and neglecting
all terms proportional to masses of the standard model particles.

\begin{table}[th]
\section*{Appendix D }
\begin{center}
\begin{tabular}{||l||r|r|r||r|r|r||}
\hline \hline  &\multicolumn{3}{|c||}{Low $\tb$} &
\multicolumn{3}{|c||}{High $\tb$} \\ \hline
Particle & M ($GeV$) & $\Sigma_1~(GeV^2)$ & $\Sigma_2~(GeV^2)$
& M ($GeV$) & $\Sigma_1~(GeV^2)$ & $\Sigma_2~(GeV^2)$ \\
\hline \hline
$\tilde{e}_L$          & 279 &    --46 &      46 & 604 &   --477 & 477 \\
\hline
$\tilde{e}_R$          & 228 &    --18 &      18 & 602 &   --408 & 408 \\
\hline
$\tilde{\nu_e}_L$      & 272 &      79 &    --79 & 599 &     866 & --867 \\
\hline
$\tilde{u}_L$          & 630 &    2802 &  --2802 & 620 &    2020 & --2020 \\
\hline
$\tilde{u}_R$          & 609 &    1158 &  --1158 & 619 &     903 &   --903 \\
\hline
$\tilde{d}_L$          & 633 &  --3460 &    3460 & 625 &  --2526 &    2526 \\
\hline
$\tilde{d}_R$          & 607 &   --575 &     575 & 621 &   --454 &     454 \\
\hline
$\tilde\mu_L$          & 279 &    --46 &      46 & 604 &   --476 &     476 \\
\hline
$\tilde\mu_R$          & 228 &    --18 &      18 & 602 &   --408 &     408 \\
\hline
$\tilde{\nu_{\mu}}_L$  & 272 &      79 &    --79 & 599 &     866 &   --866 \\
\hline
$\tilde{c}_L$          & 630 &    2802 &  --2802 & 620 &    2020 &  --2020 \\
\hline
$\tilde{c}_R$          & 609 &    1158 &  --1158 & 619 &     903 &   --903 \\
\hline
$\tilde{s}_L$          & 633 &  --3460 &    3460 & 625 &  --2526 &    2526 \\
\hline
$\tilde{s}_R$          & 607 &   --575 &     575 & 621 &   --454 &     454 \\
\hline
$\tilde{\nu_{\tau}}_L$ & 272 &      79 & --79 & 521 &     582 &   --582 \\
\hline
$\tilde\tau_1$         & 228 &    --17 &      17 & 431 &     217 &      39 \\
\hline
$\tilde\tau_2$         & 279 &    --47 &      48 & 528 &     954 &     535 \\
\hline
$\tilde{t}_1$          & 479 &   18905 &    8266 & 372 &   82031 &    2243 \\
\hline
$\tilde{t}_2$          & 586 & --29357 &   43024 & 430 &--133126 &    9441 \\
\hline
$\tilde{b}_1$          & 563 &  --2449 &    2533 & 348 &   96751 &  --1480 \\
\hline
$\tilde{b}_2$          & 607 &   --559 &     592 & 425 &--158449 &    3632 \\
\hline
$\tilde\chi^{\pm}_1$   & 231 &   --206 &    --17 &  45 &     345 &     --2 \\
\hline
$\tilde\chi^{\pm}_2$   & 552 &    1014 &    3717 & 188 &   --909 &      84 \\
\hline
$\tilde\chi^0_1$       &--553 &  --2460 &  --1375 & --166  & --258 &     --6
\\ \hline
$\tilde\chi^0_2$       & 116 &      64 &      14 &  26 &    2034 &     --7 \\
\hline
$\tilde\chi^0_3$       & 231 &    --49 &     --4 &  46 &      55 &       0 \\
\hline
$\tilde\chi^0_4$       & 545 &    2202 &  --1241 & 180 &    1012 &    --27 \\
\hline
$h$                    &  82 &    10 &  --6 & 107 &   35 &   --7 \\ \hline
$H$                    & 704 & 3167 &  294 & 180 &  63 &   --1 \\ \hline
$A$                    & 710 &    0 &    0 & 180 &    0 &     0 \\ \hline
$H^{\pm}$              & 711 &   2100 &    2100 & 199 & 19 & 19 \\ \hline
$W^{\pm}$              &  80 & --54 & --54 &  80 & --54 & --54  \\ \hline
$Z^0$                  &  92 &    0 &    0 & 92 & 0 & 0 \\ \hline
$\tau$                 &  1.7 &    0 &    0 &   1.7 & 0 & 0 \\ \hline
$t$                    & 170 &    0 & --342 & 170 & 0 & --261 \\ \hline
$b$                    & 4.3 & 0 & 0 &   4.3 & 7 &  0  \\ \hline \hline
Total                 & & --8073 & 58337& &--108842 & 13716 \\ \hline \hline
\end{tabular}
\caption{ \label{t2} Numerical values for the
one-loop contributions to the Higgs potential $\Sigma$'s for the low
and high $\tan\beta$ cases.}
\end{center}
\end{table}

%\clearpage
%\setcounter{table}{2}
\begin{table}
\begin{center}
\begin{tabular}{||l||r|r|r||}
\hline \hline
 & $\Delta_{11}$ & $\Delta_{12}$ & $\Delta_{22}$ \\
\hline \hline
Top-stop        &  72821 & --40459 &  34672 \\  \hline
Bottom-sbottom  &    341 &   --225 &    190 \\  \hline
Tau-stau        &      5 &     --7 &     12 \\  \hline
Chargino        & --6861 &    3713 & --2804 \\  \hline
Neutralino      & --38512 &    1923 & --1464 \\  \hline
Neutral Higgses & --5293 &    3037 & --1615 \\  \hline
Charged Higgses &     33 &      59 &    106 \\  \hline
Gauge bosons    &    --6 &    --12 &   --21 \\  \hline
1/2 generations &    131 &   --160 &    347 \\  \hline
Tau-sneutrino   &      8 &      13 &     24 \\  \hline
\hline
Total           &  57328 & --32144 &  29448 \\  \hline
\hline
\end{tabular}
\end{center}
\caption{\label{t3} $\Delta$'s for the low $\tan\beta$ case.}
\end{table}
%\samepage
\begin{table}
\begin{center}
\begin{tabular}{||l||r|r|r||}
\hline \hline
 & $\Delta_{11}$ & $\Delta_{12}$ & $\Delta_{22}$ \\
\hline \hline
Top-stop        & 103847 & --2441 &  5925 \\  \hline
Bottom-sbottom  & 150840 & --3198 &   204 \\  \hline
Tau-stau        &  --251 &      7 &    26 \\  \hline
Chargino        & --2487 &    103 & --420 \\  \hline
Neutralino      & --1642 &     70 & --163 \\  \hline
Neutral Higgses &    --3 &      5 &     7 \\  \hline
Charged Higgses &      0 &      2 &    97 \\  \hline
Gauge bosons    &      0 &    --1 &  --27 \\  \hline
1/2 generations &      0 &     98 &   490 \\  \hline
Tau-sneutrino   &      0 &    --1 &    50 \\  \hline
\hline
Total           & 250305 & --5357 &  6189 \\  \hline
\hline
\end{tabular}
\end{center}
\caption{\label{t4} $\Delta$'s for the large $\tan\beta$ case.}
\end{table}

\end{document}